# Uncovering the Effects of Metal Contacts on Monolayer $MoS_2$

**Kirstin Schauble,[1] Dante Zakhidov,[2] Eilam Yalon,[1,3] Sanchit Deshmukh,[1] Ryan W. Grady,[1] Kayla A. Cooley,[4] Connor J. McClellan,[1] Sam Vaziri,[1] Donata Passarello,[5] Suzanne E. Mohney,[4] Michael F. Toney,[5] A.K. Sood,[6] Alberto Salleo,[2] and Eric Pop[1,2,7,*]**

[1]Department of Electrical Engineering, Stanford Univ., Stanford, CA 94305, USA. [*]epop@stanford.edu
[2]Department of Materials Science and Engineering, Stanford Univ., Stanford, CA 94305, USA
[3]Present address: Andrew and Erna Viterbi Department of Electrical Engineering, Technion 3200003, Israel
[4]Department of Materials Science and Engineering, The Pennsylvania State Univ., University Park, PA 16802, USA
[5]Stanford Synchrotron Radiation Lightsource, SLAC National Accelerator Laboratory, Menlo Park, USA
[6]Department of Physics, India Institute of Science, Bangalore 560012, India.
[7]Precourt Institute for Energy, Stanford Univ., Stanford, CA 94305, USA

**ABSTRACT: Metal contacts are a key limiter to the electronic performance of two-dimensional (2D) semiconductor devices. Here we present a comprehensive study of contact interfaces between seven metals (Y, Sc, Ag, Al, Ti, Au, Ni, with work functions from 3.1 to 5.2 eV) and monolayer $MoS_2$ grown by chemical vapor deposition. We evaporate thin metal films onto $MoS_2$ and study the interfaces by Raman spectroscopy, X-ray photoelectron spectroscopy, X-ray diffraction, transmission electron microscopy, and electrical characterization. We uncover that, 1) ultrathin oxidized Al dopes $MoS_2$ *n*-type (>$2\times10^{12}$ $cm^{-2}$) without degrading its mobility, 2) Ag, Au, and Ni deposition causes varying levels of damage to $MoS_2$ (broadening Raman E' peak from <3 $cm^{-1}$ to >6 $cm^{-1}$), and 3) Ti, Sc, and Y react with $MoS_2$. Reactive metals must be avoided in contacts to monolayer $MoS_2$, but control studies reveal the reaction is mostly limited to the top layer of multilayer films. Finally, we find that 4) thin metals do not significantly strain $MoS_2$, as confirmed by X-ray diffraction. These are important findings for metal contacts to $MoS_2$, and broadly applicable to many other 2D semiconductors.**

As nanoscale electronic devices decrease in size, or as new materials are introduced, the role of their metal contacts becomes increasingly important. For example, contacts to transistors based on organic semiconductors, which are hindered by metal reactivity and poor band alignment,[1, 2] are the largest hurdle for realizing high-frequency operation.[3] Even traditional semiconductors like silicon and germanium suffer from Fermi level pinning at the contacts, which creates an energy barrier for carrier injection, leading to parasitic contact resistance.[4, 5] As device dimensions and contact areas continue to shrink, these parasitic resistances increasingly limit transistor performance by dominating the total device resistance.[6]

Among new materials, two-dimensional (2D) semiconductors, such as $MoS_2$, show promise toward extreme miniaturization of electronics due to superior electrical properties in atomically thin channels compared to silicon-on-insulator (SOI).[7] However, 2D devices also suffer from Fermi level pinning at the metal interface, potentially causing large Schottky barriers (and contact resistance) both for electron and hole injection.[8, 9] Efforts have been made to de-pin the Fermi level and tune the Schottky barrier height of contacts to 2D materials by stamp-transferring metal contacts[10] and by transferring hexagonal boron nitride (*h*-BN) as an interlayer,[11] but these approaches are not industrially scalable and have not yet demonstrated improved contact resistance because they introduce an additional van der Waals tunneling resistance. The lowest contact resistance to undoped monolayer $MoS_2$ (~1 kΩ·µm) is currently achieved using electron beam (e-beam) evaporated Au or Ag.[12-14] However, further reduction of contact resistance by an order of magnitude is necessary for sub-10 nm transistors.[7]

Improving contact resistance to 2D materials requires better understanding of the metal-2D interface, which remains limited in its scope today. Previous studies used X-ray photoelectron spectroscopy (XPS) or cross-section transmission electron microscopy (TEM) to show that reactions between metals and *multilayer* $MoS_2$ can occur.[15-19] However, it is unclear how many layers deep the reactions penetrate the 2D material, which is important for contacts to monolayer vs. multilayer materials. A previous study points to the lack of interfacial reaction as a requirement for epitaxy of metals on 2D materials,[20] and Ag, Au, Pb, Pd, Pt, Al, Cu, and Zn films were found (by TEM) to grow epitaxially on $MoS_2$, indicating that these metals do not react with $MoS_2$.[21] Other studies have reported large strain effects from thin (1-3 nm) evaporated metals on $MoS_2$ using Raman spectroscopy.[22-25] However, Raman analysis provides only an indirect measurement of strain by assuming shifts in $MoS_2$ Raman peaks are solely due to strain, so quantification using a direct lattice constant measurement technique, such as X-ray diffraction (XRD), remains necessary.

Here we conduct the first comprehensive study of contact interfaces between several metals and monolayer $MoS_2$ by Raman spectroscopy, XPS, grazing incidence XRD, plan view TEM, and electrical characterization. We deposit thin metal films (Y, Sc, Ag, Al, Ti, Au, and Ni) using high-vacuum e-beam evaporation (~$10^{-7}$ Torr) onto monolayer $MoS_2$ (additional details in Methods). Raman spectroscopy is used to characterize the metal-coated $MoS_2$ samples and provide insight on changes in $MoS_2$ carrier concentration,[26] defects,[27] grain size,[28] and strain.[29] Compared to other techniques, Raman spectroscopy can be done quickly, non-destructively, and with sub-micron spatial resolution, which makes it easy to distinguish between $MoS_2$ monolayers and bilayers, for example. Various techniques are then used to further refine the observations from Raman characterization: 1) XPS reveals details about chemical bonds and reactions, 2) XRD measures lattice spacings and therefore provides direct strain information, and 3) plan view TEM displays the metal morphology on the 2D material. Finally, 4) we build transistor test structures allowing us to correlate the observations from these analytical techniques with changes in electrical transport characteristics (e.g. mobility and doping) of the $MoS_2$ beneath the metal film.

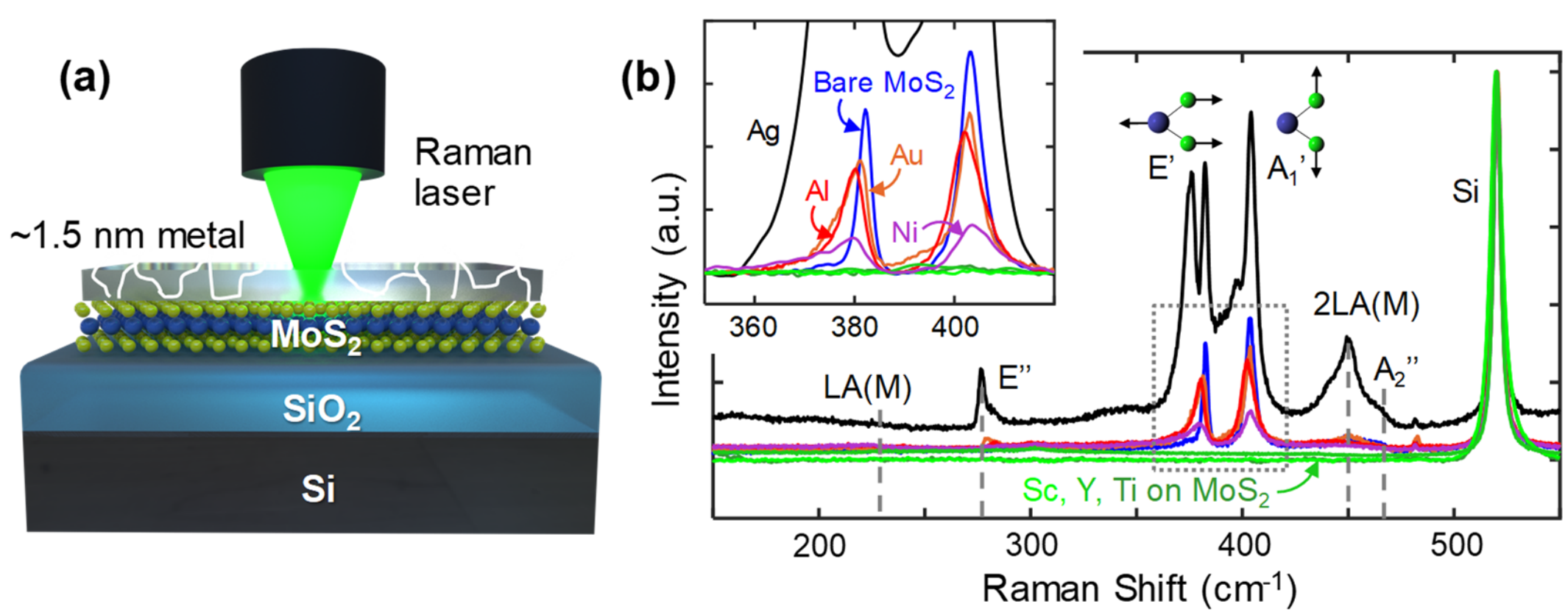

**Figure 1:** (a) Raman measurement schematic of monolayer $MoS_2$ coated with ~1.5 nm of evaporated metals. White lines illustrate grain boundaries of the discontinuous metal film (note some metals such as Au and Ag are much more discontinuous). (b) Raman spectra of monolayer $MoS_2$ bare (blue) and coated with thin metal films. All $MoS_2$ and substrate (Si) peaks are labeled. Atomic cartoons represent in-plane (E' peak) and out-of-plane ($A_1$' peak) vibrations of $MoS_2$. The inset shows a zoomed in view of the $MoS_2$ E' and $A_1$' peaks (from the boxed region, with baseline subtracted). Changes in $MoS_2$ peaks indicate modifications of $MoS_2$ due to metal evaporation. The $MoS_2$ Raman signal is no longer detectable after deposition of Sc, Y, and Ti due to interfacial reactions with $MoS_2$. Al dopes $MoS_2$ *n*-type, causing a red-shift in the $A_1$' peak. Ag, Au and Ni damage $MoS_2$, evidenced by broadening of the E' peak. Ag enhances the $MoS_2$ Raman signal due to a surface enhanced Raman scattering (SERS) effect with the 532 nm laser, as reported previously.[24, 25] This SERS effect also results in the presence of additional $MoS_2$ peaks, i.e. E'' peak at ~287 $cm^{-1}$, 2LA(M) peak at ~452 $cm^{-1}$, and $A_2$'' peak at ~465 $cm^{-1}$.[28, 30]

| | $\Phi_M$ (eV) | E' peak ($cm^{-1}$) | $A_1$' peak ($cm^{-1}$) | E' FWHM ($cm^{-1}$) | Strain ε or Δε (XRD) | Doping | Reactions (XPS) |
|---|---|---|---|---|---|---|---|
| **Bare $MoS_2$** | -- | 382.8 | 404.0 | 2.8 | ε = 0.4-0.5% | -- | -- |
| **Ni on $MoS_2$** | 5.15 | 380.4 | 404.4 | 6.3 | Δε = 0.037% | -- | $NiO_x$ |
| **Au on $MoS_2$** | 5.1 | 378.6 | 403.6 | 3.6 | Δε = 0.012% | -- | -- |
| **Ti on $MoS_2$** | 4.33 | -- | -- | -- | Δε = -- | -- | $TiO_x + Ti_xS_y$ |
| **Al on $MoS_2$** | 4.28 | 380.2 | 402.9 | 3.5 | Δε = 0.014% | n-type | $AlO_x$ |
| **Ag on $MoS_2$** | 4.26 | 375.1 | 404.0 | 3.4 | Δε = 0.003% | -- | -- |
| **Sc on $MoS_2$** | 3.5 | -- | -- | -- | -- | -- | $ScO_x + Sc_xS_y$ |
| **Y on $MoS_2$** | 3.1 | -- | -- | -- | -- | -- | $YO_x + Y_xS_y$ |

**Table 1:** Summary of work functions ($\Phi_M$ for polycrystalline metals[31, 32]), extracted $MoS_2$ Raman peak fitting, and effects of metals on $MoS_2$. Work function of $MoS_2$ is not listed, as it can vary from ~4 eV (*n*-type, e.g. after $AlO_x$ doping in this work) to ~6 eV (*p*-type), depending on doping. Tensile strain values reported are those measured by grazing incidence XRD. For bare $MoS_2$, strain (ε) is calculated by comparing the lattice spacing of as-grown $MoS_2$ with that of transferred $MoS_2$, where the range is due to variability between growths (see Supporting Information Section 6). Strain values for metal-coated $MoS_2$ are listed as Δε, with respect to the as-grown bare $MoS_2$. Doping is based on $A_1$' Raman peak shifting and $V_T$ shift in electrical characterization, with Al showing the most evident doping effects. Reaction products listed are based on XPS measurements (see Supporting Information Section 2).

## RESULTS AND DISCUSSION

Figure 1a displays the schematic for the Raman measurement of metal-coated monolayer $MoS_2$. The $MoS_2$ was grown by chemical vapor deposition (CVD)[33, 34] directly on a 90 nm $SiO_2$/Si substrate. The nominal deposited metal thickness (~1.5 nm) is transparent to the Raman laser, allowing straightforward measurement of the underlying $MoS_2$ Raman signatures. Figure 1b shows the Raman spectra of monolayer $MoS_2$ bare and coated with the ultrathin metal films, where the E' peak corresponds to in-plane and the $A_1$' peak to out-of-plane atomic vibrations, and intensities are normalized by the Si peak. The inset displays a zoomed in view of changes in the $MoS_2$ E' and $A_1$' peaks after metal deposition, which will be discussed in detail throughout this paper. For each metal on $MoS_2$, three samples were prepared and five spots on each sample were measured, where Figure 1b illustrates representative data and Table 1 shows the average extracted peak information (see Supporting Information Section 1 for details on peak fitting).

**Interfacial Reactions.** We first discuss three of the low work function metals (Y, Sc and Ti: $\Phi_M$ ~ 3.1 to 4.3 eV), which could be used for *n*-type contacts to 2D semiconductors given the better theoretical alignment of their Fermi level with the 2D material conduction band.[35] Figure 1b reveals that when monolayer $MoS_2$ is coated with ultrathin films of these metals, the $MoS_2$ Raman signal is no longer detectable. A broader spectral range (50 to 1500 $cm^{-1}$) does not show additional Raman peaks, except those related to the Si substrate (see Supporting Information Section 1). This suggests that interfacial reactions between these metals and $MoS_2$ occur and the resulting compounds do not have Raman-active modes. However, the Raman signal remains present for Y, Sc, and Ti-coated *bilayer* $MoS_2$ indicating that the reactions mostly affect the top layer of $MoS_2$ (Supporting Information Figure S1). This is also an important finding, because bilayer and multilayer $MoS_2$ may have lower Schottky barrier for electron injection than monolayer,[35, 36] and thus could benefit from contacts with these reactive metals (provided they do not oxidize), similar to the process of contact silicidation in mainstream silicon technology.[37]

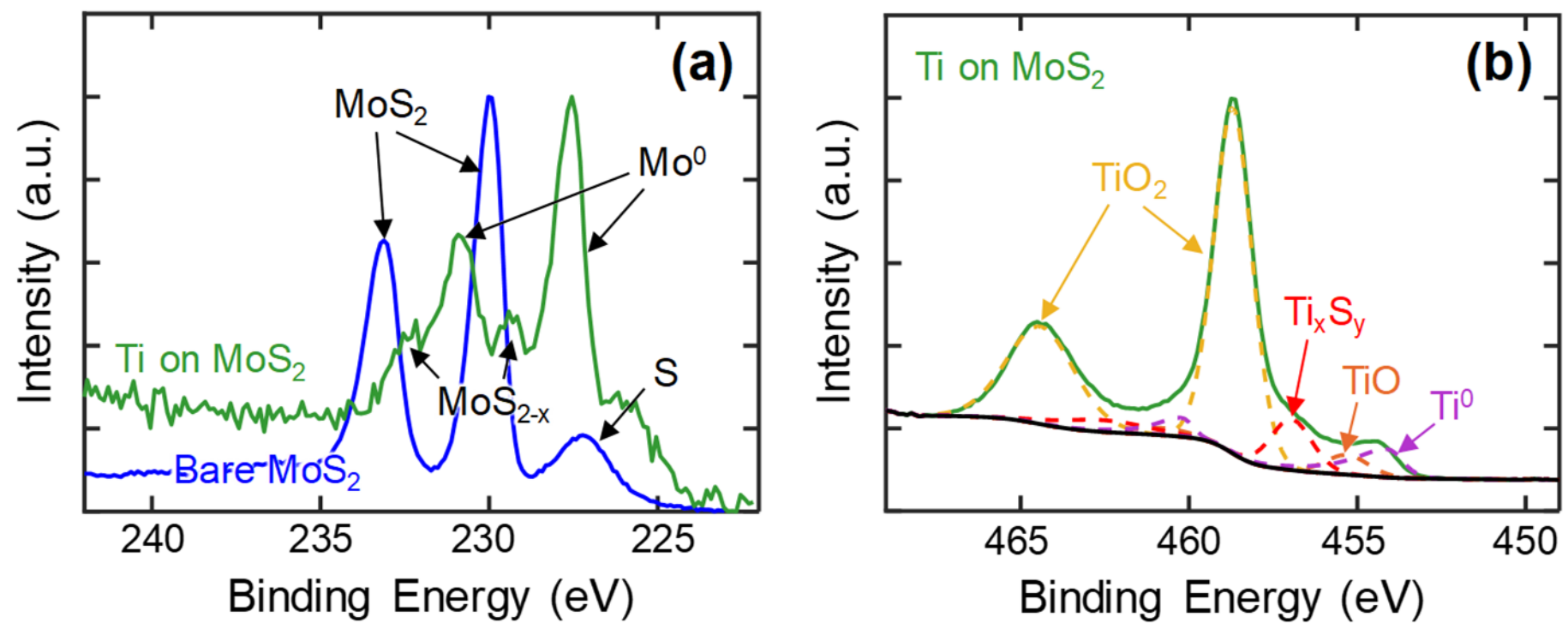

**Figure 2:** (a) XPS Mo 3d spectrum of $MoS_2$ bare (blue) and coated with 1.5 nm Ti (green), normalized by their tallest peaks. These spectra reveal evidence of metallic Mo and sulfur-deficient $MoS_{2-x}$ as byproducts of the reaction between Ti and $MoS_2$. (b) XPS Ti 2p spectrum of Ti-coated $MoS_2$ (green) and doublet peak fits (dashed lines). Note that $Ti_xS_y$, TiO, and $Ti^0$ doublet peaks at higher energy are included in the fit, but the arrows are omitted in this figure. These spectra show evidence of oxidized Ti, $Ti_xS_y$, and metallic Ti, verifying reaction at Ti-$MoS_2$ interface. XPS spectra for Sc and Y also show similar signs of oxidation and reaction with $MoS_2$ (see Supporting Information Figure S3).

Next, we took XPS data to better understand these interfacial reactions. Figure 2a displays the Mo 3d XPS spectra, showing reaction byproducts of metallic Mo and sulfur-deficient $MoS_{2-x}$ after Ti deposition, confirming that Ti chemically bonds with the sulfur from $MoS_2$. Figure 2b shows the XPS Ti 2p spectrum, which reveals prominent Ti oxidation and a $Ti_xS_y$ shoulder. Note that there may be additional Ti oxidation states between the $TiO_2$ and metallic Ti peaks, thus the fits shown here are a simplified approximation. Following a study by Freedy *et al.*[38] on the oxide composition of Ti based on deposition rate and pressure, the Ti in our study (deposited at ~0.5 Å/s and ~$10^{-7}$ Torr) is only partially oxidized (between 25 and 67%) during deposition. This allows the unoxidized Ti to react with $MoS_2$ during deposition.

Similar oxidation and reactions with $MoS_2$ occur for Y and Sc, but none of the other metals in this study (Al, Ag, Ni, or Au) show evidence of reactions with $MoS_2$. XPS data for all samples are shown in Supporting Information Section 2, where the ultrathin Al and Ni are found to oxidize after exposure to air, while Au and Ag remain unreactive with $MoS_2$ and oxygen. However, the Ag-coated $MoS_2$ sample shows a broadened S 2p spectrum compared to bare $MoS_2$. Since the Mo 3d spectrum remains unaffected, this suggests that Ag is tarnishing in air (and not reacting with $MoS_2$). To validate this, we performed selective area electron diffraction (SAED) on the Ag-coated $MoS_2$ sample, which confirms the presence of $MoS_2$, polycrystalline Ag, and $Ag_2S$ (see Supporting Information Section 3 for SAED data and Methods for SAED sample preparation). Table 1 shows a summary of reactions for each sample studied.

The observed reactions between metals and $MoS_2$ mostly match expectations based on thermodynamic enthalpies of formation ($\Delta H$) for each metal sulfide and oxide (values reported in Supporting Information Section 4). For example, Y, Sc, and Ti sulfides have $\Delta H$ (per mole of solid S) that are much more negative than $MoS_2$, meaning the sulfur atoms in $MoS_2$ prefer to bond with Y, Sc, and Ti than Mo, assuming small entropies of formation and no kinetic limitations. Ag, Au, and Ni have less negative $\Delta H$ with sulfur and were not experimentally found to react with $MoS_2$. Indeed, a previous report has predicted Au to be in thermodynamic equilibrium with $MoS_2$, while ternary phases in the Ag-Mo-S and Ni-Mo-S systems are reported without corresponding thermodynamic data.[39] The only exception is Al, which has $\Delta H$ favoring reaction with sulfur but is not experimentally found to react with $MoS_2$. This has also been observed in previous studies,[15, 17] and can be explained by a larger kinetic barrier for reaction with sulfur, or by Al oxidation since $Al_2O_3$ also has a very negative $\Delta H$ (see Supporting Information Section 4).

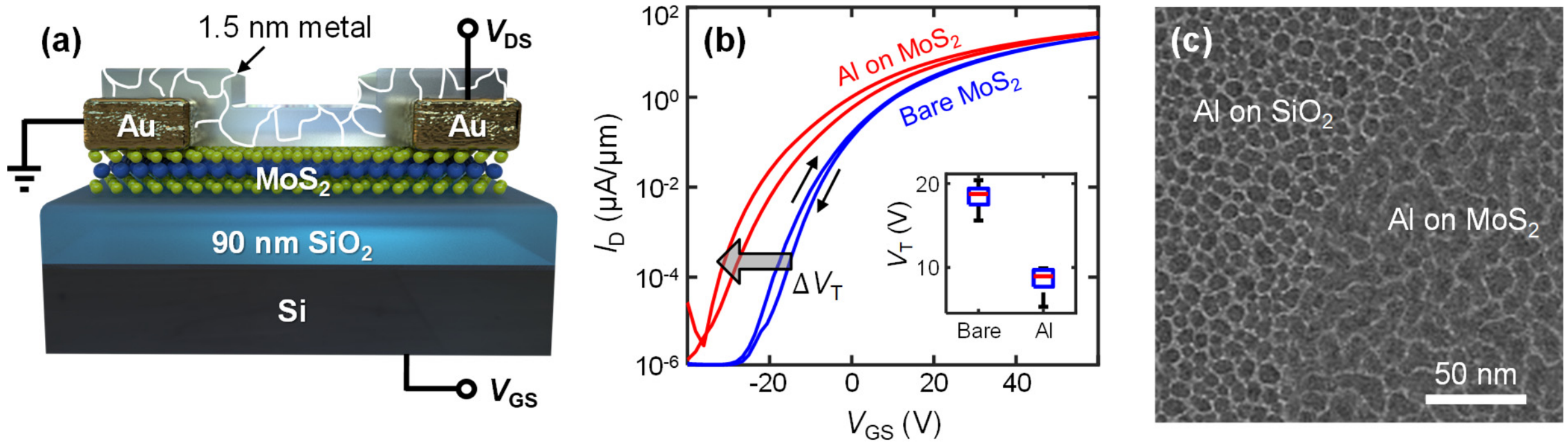

**Figure 3:** (a) Schematic of back-gated monolayer $MoS_2$ transistors coated with thin metals (which are neither continuous nor conductive). (b) Device characteristics ($L$ = 3 μm, $V_{DS}$ = 1 V) before and after capping with 1.5 nm oxidized Al. An average threshold voltage shift $\Delta V_T = -10.1$ V is seen, signifying $n$-type doping of 2.4 ± 0.5 × $10^{12}$ electrons/cm$^2$. Small arrows denote forward and backward voltage sweeps, revealing minimal hysteresis. Inset shows box plot of $V_T$ for bare and Al-coated $MoS_2$ transistors (10 devices each) for gate lengths of 3 to 6 μm. (c) Plan view TEM of monolayer CVD $MoS_2$ covered with ~1.5 nm Al on an $SiO_2$ TEM grid, showing Al grains are discontinuous.

**Doping.** Figure 1b reveals red-shifting of the $A_1$' Raman peak in Al-coated $MoS_2$, which has been shown to correlate with increased electron concentration (4.5 ± 0.5 × $10^{12}$ electrons/cm$^2$ per cm$^{-1}$ peak shift).[26, 40] Ultrathin Al oxidizes into sub-stoichiometric $AlO_x$, which is known to behave as an electron charge transfer doping layer for $MoS_2$.[41-43] Based on the observed $A_1$' peak shift, the ~1.5 nm oxidized Al contributes 4.8 ± 2.7 × $10^{12}$ electrons/cm$^2$ doping of $MoS_2$, where the uncertainty comes from variation in $A_1$' peak position for bare vs. Al-coated $MoS_2$. Electron doping pushes the Fermi level near the conduction band, which can cause optical phonons to interact with the continuum of electronic states. This is evidenced in the $MoS_2$ Raman spectrum, where an asymmetric Fano line shape can be fit to the Al-coated $MoS_2$ E' Raman peak. This Fano line shape has also been reported as evidence of doping for other semiconductors such as Si, Ge, and carbon nanotubes.[44-47] See Table 1 for positions of the $A_1$' peak for each metal on $MoS_2$, and Supporting Information Section 1 for further discussion of the Fano fits for each metal on $MoS_2$.

To verify the doping effect inferred from Raman analysis, we also fabricated $MoS_2$ transistors, with $SiO_2$/Si back-gates and Au contacts, then deposited the same ultrathin ~1.5 nm of metals on the device channels (Figure 3a). The metals are discontinuous, not shorting the source and drain, and preserving an $MoS_2$ transistor on/off ratio >$10^3$ for the non-reactive metals, as shown in Figures 3b and 4a. (Control structures employing the same ultrathin metal layer on $SiO_2$ without the $MoS_2$ show no measurable conduction.) Figures 3c and 4b display plan view TEM images of the thin, discontinuous Al and Au on $MoS_2$, using the same TEM grids from the SAED analysis. These images further confirm that the $MoS_2$ device source and drain are not shorted by the thin metal. Supporting Information Figure S5 shows additional plan view TEM images of metal-coated $MoS_2$ samples, including Ag and Ni.

Figure 3b displays measured drain current vs. back-gate voltage of the monolayer $MoS_2$ transistors before and after capping with Al, revealing an average threshold voltage shift $\Delta V_T = -10.1$ V among 10 devices (see inset of Figure 3b), while preserving mobility. This corresponds to $n$-type doping of 2.4 ± 0.5 × $10^{12}$ electrons/cm$^2$ estimated using $\Delta n = |\Delta V_T| C_{ox}/q$, where $\Delta n$ is the change in electron concentration, $C_{ox}$ ≈ 38 nF/cm$^2$ is the oxide capacitance, and $q$ is the elementary charge. This is consistent, within error bars, with the change in carrier concentration estimated independently from Raman analysis, and both support the observation that $AlO_x$ $n$-type dopes $MoS_2$. Thus, Al is not a good contact metal (as $AlO_x$ is insulating), but can be used as a damage-free and tunable (e.g. by varying $AlO_x$ thickness)[43] dopant of $MoS_2$.

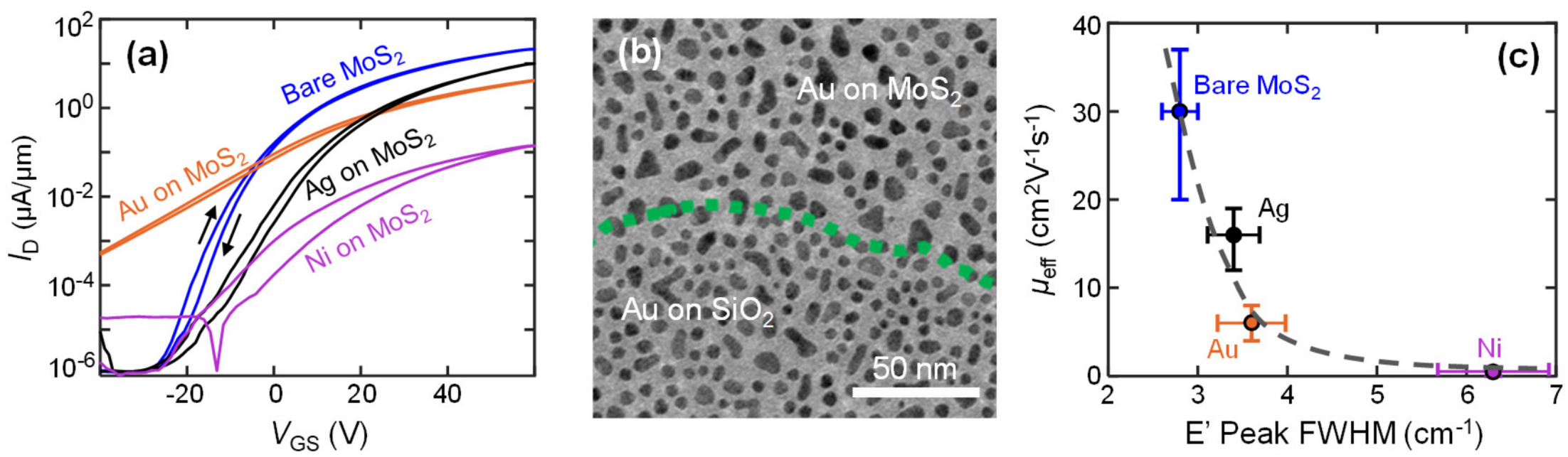

**Figure 4:** (a) Monolayer $MoS_2$ device transfer characteristics ($L = 3$ μm, $V_{DS} = 1$ V) before and after capping with 1.5 nm Ag, Au, and Ni result in degraded on/off ratio and mobility. Small arrows denote forward and backward voltage sweeps. (b) Plan view TEM of $MoS_2$ covered with 1.5 nm Au shows that Au grows in discontinuous islands on both $MoS_2$ and $SiO_2$ (green dashed line shows border of $MoS_2$). (c) Electron mobility (at $n \approx 7 \times 10^{12}$ cm$^{-2}$) vs. $MoS_2$ E' Raman peak FWHM shows that larger E' FWHM correlates with lower mobility (gray dashed line to guide the eye).

**Damage.** We recall that Figure 1b revealed significant broadening of the $MoS_2$ E' Raman peaks after deposition of ultrathin Ni, Au, and Ag. Such broadening is known to occur due to phonon confinement by reduced grain sizes or increased disorder within $MoS_2$.[28, 48-52] Thus, the E' peak broadening is an indication that the $MoS_2$ is damaged during the evaporation of these metals, and Table 1 quantifies the full width at half maximum (FWHM) of this peak for monolayer $MoS_2$ coated with each metal (see Supporting Information Section 1 for details on peak fitting). We note that possible $MoS_2$ damage due to Au evaporation has also been reported previously using cross-sectional TEM, showing defects and atomic disorder in multilayer $MoS_2$.[10] Although we see evidence of disorder in $MoS_2$ after metal deposition via E' peak broadening, we do not see the appearance of an LA(M) peak (Figure 1b), which has been shown to appear in very defective $MoS_2$ (with inter-defect distance $L_D < 3.2$ nm or defect density $> 10^{13}$ cm$^{-2}$).[48, 50] Therefore, even the deposition of metals which here induce the most defects in $MoS_2$ (Ni, Au, Ag) do not cause defect densities high enough for the emergence of the LA(M) peak.

Figure 4a shows electrical characteristics of monolayer $MoS_2$ channels covered with ultrathin Ni, Au, and Ag, while Figure 4b displays the discontinuous Au on $MoS_2$. These results show degradation of the transistor on/off ratio from $>10^7$ for bare $MoS_2$ to $<10^4$ for Ni- and Au-coated $MoS_2$. The effective mobility (at carrier density $n \approx 7 \times 10^{12}$ cm$^{-2}$) extracted using transfer length method (TLM) structures,[53, 54] also degrades after evaporation of these metals. Figure 4c shows a clear trend between decreasing effective mobility and increased broadening of the $MoS_2$ E' Raman peak. The level of $MoS_2$ damage (increasing E' FWHM and decreasing mobility) also scales approximately with metal melting temperature. These results indicate that Raman spectroscopy can be a useful tool for estimating the electrical quality of an $MoS_2$ film. See Supporting Information Section 5 for additional details on TLM mobility calculations and electrical characteristics of $MoS_2$ channels covered with the reacting metals (Y, Sc, and Ti).

There are potential trade-offs of inducing damage to semiconductor contact regions. We have shown that increasing defects in $MoS_2$ lowers its effective mobility, so lateral conduction under the contacts will be hindered. However, defects can also improve the contact resistance between the metal and $MoS_2$, as conductive atomic force microscopy (C-AFM) has shown that charge injection can be higher at $MoS_2$ defect sites.[55] Other experiments have found up to a 50% decrease in contact resistance when an Ar ion beam was used to induce defects in multilayer $MoS_2$ before Ni contact deposition.[56] Similarly, etching holes into graphene at the contact regions was found to create improved edge current injection.[57] While the defect density is most likely determined by the metal deposition instrument and parameters, it is interesting to note that some of the best contacts to $MoS_2$ have been obtained with evaporated Ag[13, 58] or Au,[12, 43, 54] suggesting these may be good contacts simply because they do not oxidize and their deposition creates "just enough" defects in $MoS_2$. In contrast, "defect-free" contacts (e.g. with *h*-BN depinning layers,[11] transfer-stamped Au[10] or In deposition[19, 59]) suffer from additional van der Waals tunneling resistance. Thus, defect engineering may play an important role in improving contacts to 2D materials.

**Strain in As-Grown $MoS_2$.** Strain in 2D materials is often characterized using Raman spectroscopy, by correlating observed peak shifting to strain.[29, 60-63] However, many other factors can affect the phonon modes of such 2D materials, causing shifts in Raman peaks. Therefore, it is important to confirm Raman-based strain estimates with a direct lattice constant measurement technique. Here, we report a direct measurement of strain in monolayer $MoS_2$ using grazing incidence XRD at a synchrotron radiation light source. By measuring a large polycrystalline $MoS_2$ film in grazing incidence with high energy X-rays, we measure the strain of monolayer $MoS_2$, which is compared with Raman-based strain analysis.

We first measure the built-in strain of our as-grown monolayer $MoS_2$. Due to the larger $MoS_2$ thermal coefficient of expansion compared to $SiO_2$, the high-temperature (850°C) CVD growth process results in tensile-strained $MoS_2$. This results in an E' Raman peak red-shift, consistent with previous observations.[64] Comparing the E' Raman peak position of CVD-grown vs. transferred $MoS_2$ (where transferring is assumed to relax built-in strain), we estimate that our as-grown $MoS_2$ is biaxially tensile strained ~0.4 to 0.5%, depending on the growth. This is based on a calibration of 4.5 $cm^{-1}$ E' peak shift per % biaxial strain, as reported by Li *et al*.[62] We then used grazing incidence XRD to analyze the monolayer $MoS_2$ in-plane (10) peak before and after $MoS_2$ transfer, which verified that the strain values measured from Raman peak shifting were accurate within ~0.02% strain (Supporting Information Section 6). The 0.02% mismatch is within the distribution of strains of CVD-grown $MoS_2$ calculated using a Williamson-Hall analysis of several monolayer $MoS_2$ growths (see Supporting Information Section 6). These Raman and XRD results also confirm the ~4.5 $cm^{-1}$ $MoS_2$ E' Raman peak red-shift per % biaxial tensile strain.[62] Based on these measurements, we find that Raman analysis can be useful to determine built-in strain in *bare* as-grown 2D materials, which has been previously used for $WSe_2$.[65] However, we will show in the next section that in more complicated systems, such as $MoS_2$ coated with metals, Raman analysis of $MoS_2$ strain is not accurate.

**Strain in Metal-Covered $MoS_2$.** As strain can affect the metal-2D semiconductor band alignment,[66] it is important to understand and accurately represent strain in the semiconductor contact regions. The Raman spectra of as-grown $MoS_2$ after metal deposition (Figure 1b) reveal red-shifting of the E' peak position, which is typically attributed to strain. The largest E' peak shift (7.6 $cm^{-1}$) from the already shifted as-grown bare $MoS_2$ is seen in the Ag-coated sample, which would suggest that the $MoS_2$ is biaxially tensile strained by ~1.7% from the thin metal, in addition to the ~0.5% built-in $MoS_2$ strain from CVD growth. Similar claims of $MoS_2$ strain due to thin (1-3 nm) Au and Ag based on Raman analysis have been previously reported,[22-25] but until now this remains unconfirmed using a direct strain measurement technique.

Here we perform grazing incidence XRD to directly measure the CVD-grown monolayer $MoS_2$ lattice constant with and without thin deposited Ni, Al, Au, and Ag. Testing three samples for each metal, we found that none of the metals appreciably strain the underlying $MoS_2$. Figure 5a shows one set of samples displaying that the $MoS_2$ (10) peak does not shift for any deposited thin metal film. Figure 5b shows the average calculated strain (across three samples for each metal) based on Raman and XRD analysis, where XRD reveals $< 0.04 \pm 0.02\%$ average change in strain from the as-grown monolayer $MoS_2$. This indicates that thin metals do not significantly strain monolayer $MoS_2$. Furthermore, the FWHM of the $MoS_2$ (10) peak (Figure 5a) is consistent across all bare and metal-coated $MoS_2$ samples (see Supporting Information Section 6), which reveals that there is also no change of $MoS_2$ strain *distribution* due to contact metals.

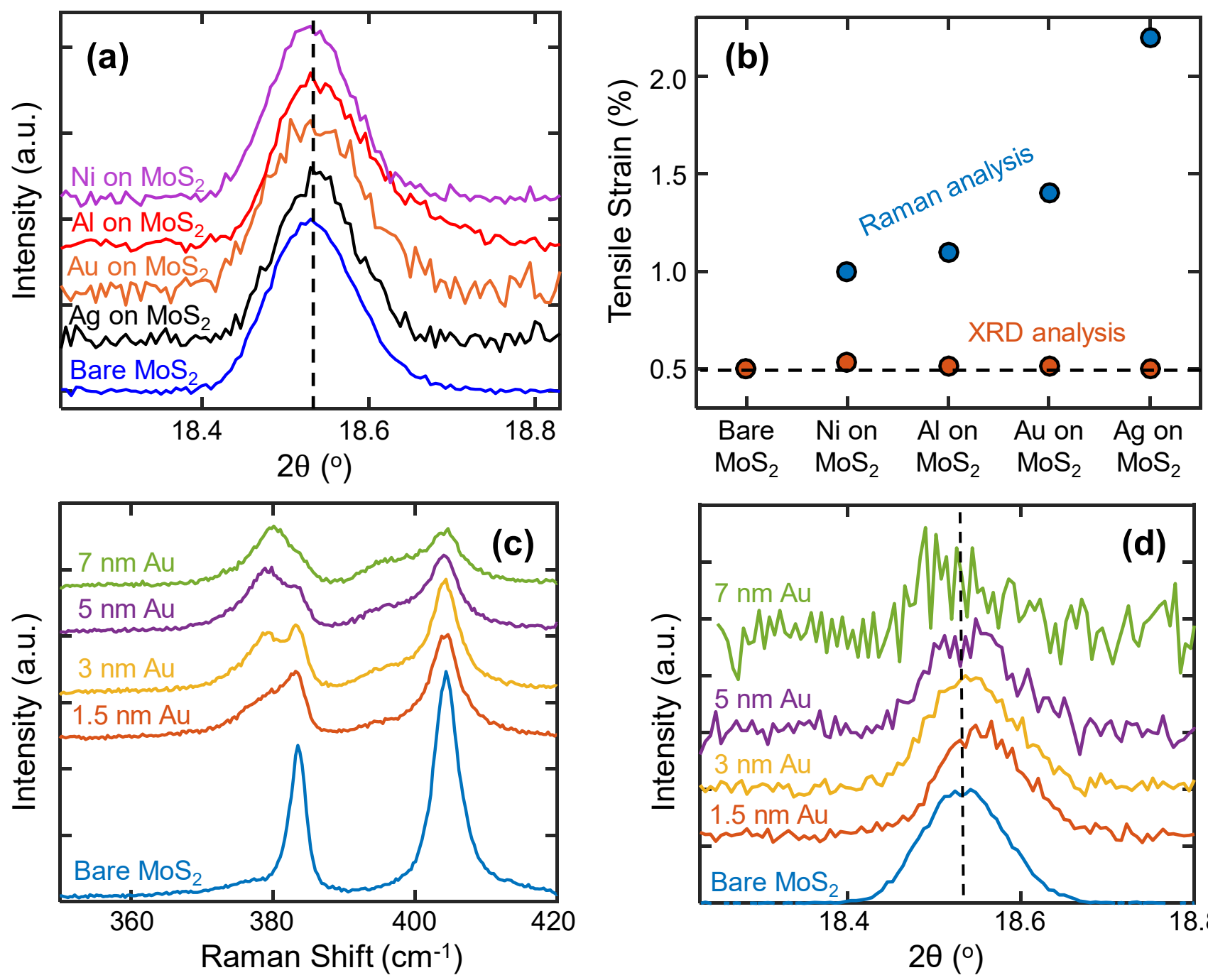

**Figure 5:** (a) Two theta values of $MoS_2$ in-plane (10) peaks with deposited contact metals show no appreciable strain compared to bare $MoS_2$. (b) Comparison of $MoS_2$ strain for each sample as characterized by both Raman and grazing incidence XRD, which reveals that Raman analysis greatly overestimates $MoS_2$ strain due to thin metals. (c) Raman spectra for $MoS_2$ bare and coated with 1.5, 3, 5, and 7 nm of deposited Au. This confirms that there are two E' peaks (since thin Au deposits in islands), and the red-shifted peak becomes more prominent as Au coverage on $MoS_2$ increases. (d) XRD peaks for the same $MoS_2$ samples coated with thicker Au confirm that even as Au gets thicker (and thus more complete coverage of Au on $MoS_2$), the measured strain of $MoS_2$ does not change.

We recall that most thin metals (especially Au, see Figure 4b) form incomplete coverage on $MoS_2$. To test if metal continuity affects the strain measurements, we deposit 1.5, 3, 5, and 7 nm of Au on monolayer $MoS_2$ and compare the Raman and grazing incidence XRD data. In Figure 5c, the Raman spectra of $MoS_2$ with thicker Au show that the shifted E' peak (from regions of $MoS_2$ covered by Au) becomes more dominant with increased Au coverage. However, grazing incidence XRD measurements on the same samples, in Figure 5d, indicate that increasing the metal thickness and coverage still does not strain the underlying $MoS_2$. Therefore, based on the XRD results presented here, using Raman peak shifts alone is likely to lead to an overestimate of strain in $MoS_2$ under thin metals.

The discrepancy between the Raman and XRD measurements of strain is due to the nature of the experimental techniques. XRD measures the constructive scattering of ordered lattice spacings, which is a direct measurement of the material strain. On the other hand, Raman-based estimates rely on calibrating shifts of phonon modes to strain, assuming that strain is the only factor affecting the phonon modes. However, shifts of the $MoS_2$ E' peak can arise from other effects, such as increased disorder (defect density),[48, 50-52] or decreased domain size[28, 49] of $MoS_2$. It has also been shown that Au and Ag-coated $MoS_2$, which exhibit the largest E' peak shifts, induce plasmon coupling with $MoS_2$ excitons.[67] This causes a coupling of phonons to the $MoS_2$ electronic continuum, which likely contributes to E' peak red-shifting.

**Discussion.** While this experiment was performed with monolayer $MoS_2$, the methodologies and results from this work can be used to draw hypotheses about contacts to other 2D materials. For example, in this study, most contact metal-$MoS_2$ reactions follow what is expected based on thermodynamics. Therefore, thermodynamic analysis can be carried out for projections of expected metal reactions with selenides and tellurides, and indeed a recent experimental report of metal epitaxy and reactivity with few-layer $WSe_2$ was consistent with calculated isothermal phase diagrams.[20] Defect sites may also enhance a 2D material's reactivity,[18, 68] which is an important consideration as selenides and tellurides tend to be more defective than sulfides, and CVD-grown 2D materials tend to be more defective than exfoliated 2D materials.[52] Additionally, since the bond energy is weaker for Mo-Te and Mo-Se than Mo-S,[52, 69] it is reasonable to expect that they would be even more susceptible to metal evaporation-induced damage and reactions at their contacts.[11] As 2D monolayers represent the ultimate channel thinness limit for all semiconductors, the lessons gleaned here could also benefit other semiconductors in this extreme limit.

Finally, we briefly discuss our vision to improve contact resistance to 2D semiconductors based on the findings in this study and lessons from past generations of contact engineering. First, Schottky barrier widths must be decreased by doping the source and drain regions. Few results, including those shown in this work with $AlO_x$, have demonstrated such doping through charge transfer from non-stoichiometric oxide capping layers to $MoS_2$.[42, 43, 58] Second, reacting the contact metal with the 2D semiconductor may be used to create a more intimate contact interface, as is well-known from silicide contacts for Si devices.[37] This intimate contact interface may also be achieved by defect engineering (tuning defect density in the contact region, e.g. by adjusting deposition techniques and parameters). However, we have shown that defects and reactions can destroy a monolayer 2D material, but do not completely destroy a bilayer material. Therefore, thicker 2D material at the contacts should be explored (e.g. by re-growth), not unlike "raised source/drain" of Si devices,[70] where only the top one or two layers react with the metal and the bottom layer(s) remain unharmed. Third, the contact metal band alignment with the 2D semiconductor may be tuned by Fermi level de-pinning,[11] or alloying and strain, as was previously achieved by alloying Ge into Si source/drain regions to improve *p*-type contacts.[71] Further efforts in each of these areas are critical to reduce contact resistance below 100 Ω·µm for 2D semiconductors.

## CONCLUSIONS

In summary, we deposited thin films of Y, Sc, Ag, Al, Ti, Au, Ni on monolayer $MoS_2$ and studied the contact interface by Raman spectroscopy, XPS, XRD, TEM, and electrical characterization. We found through Raman spectroscopy and XPS that low work function metals (Y, Sc, and Ti, often used as adhesion layers) oxidize and react with the underlying monolayer $MoS_2$, which negatively impacts the contact resistance. We also showed through Raman spectroscopy and electrical characterization that ultrathin Al oxidizes and dopes $MoS_2$, while Ag, Au, and Ni show varying levels of damage to $MoS_2$. Lastly, we analyzed the strain effects of metals on $MoS_2$ using Raman spectroscopy and XRD, noting discrepancies between the two techniques, and we conclude that the thin evaporated metals studied do not strain $MoS_2$, contrary to previous reports. The findings in this work, as well as the implications discussed for other 2D materials, are crucial for understanding and improving metal contacts to 2D semiconductor devices.

## METHODS

**Sample Preparation.** Monolayer $MoS_2$ is grown directly onto 90 nm $SiO_2$/Si substrates by CVD using solid sulfur and $MoO_3$ precursors, and perylene-3,4,9,10 tetracarboxylic acid tetrapotassium salt (PTAS) as a seeding layer. The growths take place at 760 Torr and 850°C, with an Ar flow rate of 30 sccm, and additional details were reported by Smithe *et al*.[33, 34] The growth results in a primarily monolayer film of $MoS_2$ crystallites with sparse regions (<10%) of bilayer $MoS_2$.[33] Thin films (~1.5 nm, measured by crystal monitor) of metals were deposited using the Kurt. J. Lesker PVD 75 e-beam evaporation system in high vacuum (~$10^{-7}$ Torr) at a rate of ~0.5 Å/s. Each metal was deposited on 3 separate $MoS_2$ growths to ensure repeatability of results.

**Raman Measurements.** All Raman measurements were taken using the Horiba Labram HR Evolution Raman System in the Stanford Nanofabrication Shared Facility. A green laser source of wavelength 532 nm was used with 2.5% incident laser power (0.12 mW) to avoid heating effects, and the spot size is less than 1 μm. An 1800 grooves/mm grating was used, resulting in a spectral resolution of ~0.3 $cm^{-1}$. Peak fitting details are described in Supporting Information Section 1.

**XPS Measurements.** All XPS measurements were taken in the Stanford Nanofabrication Shared Facility using a PHI VersaProbe III scanning XPS microprobe. The XPS instrument uses a monochromatized Al(Ka) radiation (1486 eV) as source of incident X-rays. We performed all measurements at a chamber pressure < $10^{-6}$ Torr. The X-ray spot size was 1400 × 100 μm with 100 W incident power (high power). Additionally, we performed auto sample neutralization to overcome sample charging effects. The XPS analysis and peak fitting was performed in CasaXPS.

**Plan View TEM Measurements.** $MoS_2$ was grown onto SPI 20 nm $SiO_2$ TEM grids, and subsequent metal evaporation was performed as described above. The $MoS_2$ growth process was the same as described in Sample Preparation, however instead of using the PTAS seeding layer directly on the growth substrate, a separate chip covered in PTAS was placed directly upstream from the growth substrate. This different growth geometry was due to the small size of the TEM grid substrates. Plan view TEM and SAED examination were performed using a FEI Talos F200X in the Materials Characterization Laboratory at The Pennsylvania State University. All samples were analyzed using an accelerating voltage of 80 kV to limit e-beam damage of the $MoS_2$. The imaging data provided insight concerning the size of metal nuclei on the $MoS_2$ surface as well as their continuity. Electron diffraction patterns provided information concerning whether the metal nuclei were epitaxial or randomly oriented on the $MoS_2$ surface.

**XRD Measurements.** All XRD measurements were taken at the Stanford Synchrotron Radiation Lightsource beam line 7-2, where the diffraction geometry is illustrated in Supporting Information Figure S11. The polycrystalline nature of the CVD-grown $MoS_2$ samples enabled the diffraction conditions necessary to obtain $MoS_2$ signal. The sample was attached to a six-circle diffractometer with the sample plane vertical during the measurements. The sample was covered with a Kapton dome and purged with helium gas to improve signal-to-noise ratio by reducing air scattering. The 14 keV (0.885 Å) incident beam was set to grazing incidence (0.1°) relative to the sample surface and the scattered radiation was collimated to 1 mrad by Soller slits and collected by a Vortex point detector. The sample was rocked during the measurement to remove potential beam damage, and each measurement was averaged over 3 exposures to further reduce noise. Based on the grazing incidence angle and sample rocking, the effective sample measurement area is ~2.8 mm wide by the whole length of the sample (~7-12 mm). Therefore, like the XPS measurements, sparse (<10%) bilayer $MoS_2$ regions are included in the XRD signal. The experimental setup was calibrated using a lanthanum hexaboride standard.

**Device Fabrication and Measurement.** All device fabrication was performed in the Stanford Nanofabrication Facility. Optical lithography is used to define probe pads, electrical contacts, and channels in three separate steps. $O_2$ plasma (10 W) is used to etch the $MoS_2$ for channel definition.[34] Au is evaporated using e-beam evaporation as a planar contact to the $MoS_2$. Finally, the substrate is loaded into a Janis vacuum probe station (~$10^{-5}$ Torr) for measurements using a Keithley 4200-SCS. After measurements, devices were coated with a thin layer (~1.5 nm) of metals using the e-beam evaporation parameters outlined above, then re-measured using the same Janis setup. At least 10 transistors were measured for each sample, before and after coating with metal, to ensure consistency of results.

## ASSOCIATED CONTENT

### Supporting Information

The Supporting Information is available free of charge on the ACS Publications website at DOI: XX.YY

Additional details on Raman measurements and analysis, XPS measurements, SAED measurements, thermodynamic analysis, Fano line fits, plan view TEM measurements, TLM mobility extraction, electrical characterization of low work function metals on $MoS_2$, strain measurements of as-grown monolayer $MoS_2$, and X-ray diffraction measurements. (PDF)

## AUTHOR INFORMATION

### Corresponding Author

*E-mail: epop@stanford.edu

### ORCID

Kirstin Schauble: 0000-0002-4130-9181

Dante Zakhidov: 0000-0003-3107-104X

Eilam Yalon: 0000-0001-7965-459X

Sanchit Deshmukh: 0000-0003-1848-2127

Ryan Grady: 0000-0002-0457-5026

Kayla Cooley: 0000-0002-6598-4296

Connor J. McClellan: 0000-0002-8733-9968

Suzanne E. Mohney: 0000-0001-5649-7640

A.K. Sood: 0000-0002-4157-361X

Eric Pop: 0000-0003-0436-8534

### Notes

The authors declare no competing financial interest.

## ACKNOWLEDGEMENTS

This work was supported in part by ASCENT, one of six centers in JUMP, a Semiconductor Research Corporation (SRC) program sponsored by DARPA. It is also supported by the National Science Foundation (NSF) EFRI 2-DARE, the Air Force grant FA9550-14-1-0251, and the Stanford SystemX Alliance. Work was performed in part at the Stanford Nanofabrication Facility and the Stanford Nano Shared Facilities, which are supported by the NSF as part of the National Nanotechnology Coordinated Infrastructure under award ECCS-1542152. Use of the Stanford Synchrotron Radiation Lightsource at SLAC National

Accelerator Laboratory, was supported by the U.S. Department of Energy, Office of Basic Energy Sciences under Contract No. DE-AC02-76SF00515. Transmission Electron Microscopy was supported by NSF through DMR 1410334. K.S. thanks the support of NSF Graduate Research Fellowship Program (GRFP) under Grant No. DGE-1656518 and Stanford Graduate Fellowships. D.Z. and R.W.G. also acknowledge support by the NSF GRFP under Grant No. DGE-1656518, and K.A.C. thanks the NSF GRFP Grant No. DGE-1255832 for support. The Knut and Alice Wallenberg Foundation partially supported S.V. through a postdoctoral fellowship. A.K.S. thanks Department of Science and Technology, India for Year of Science Professorship which funded the travel to facilitate the collaboration.

## Table of Contents (T.O.C.) Graphic

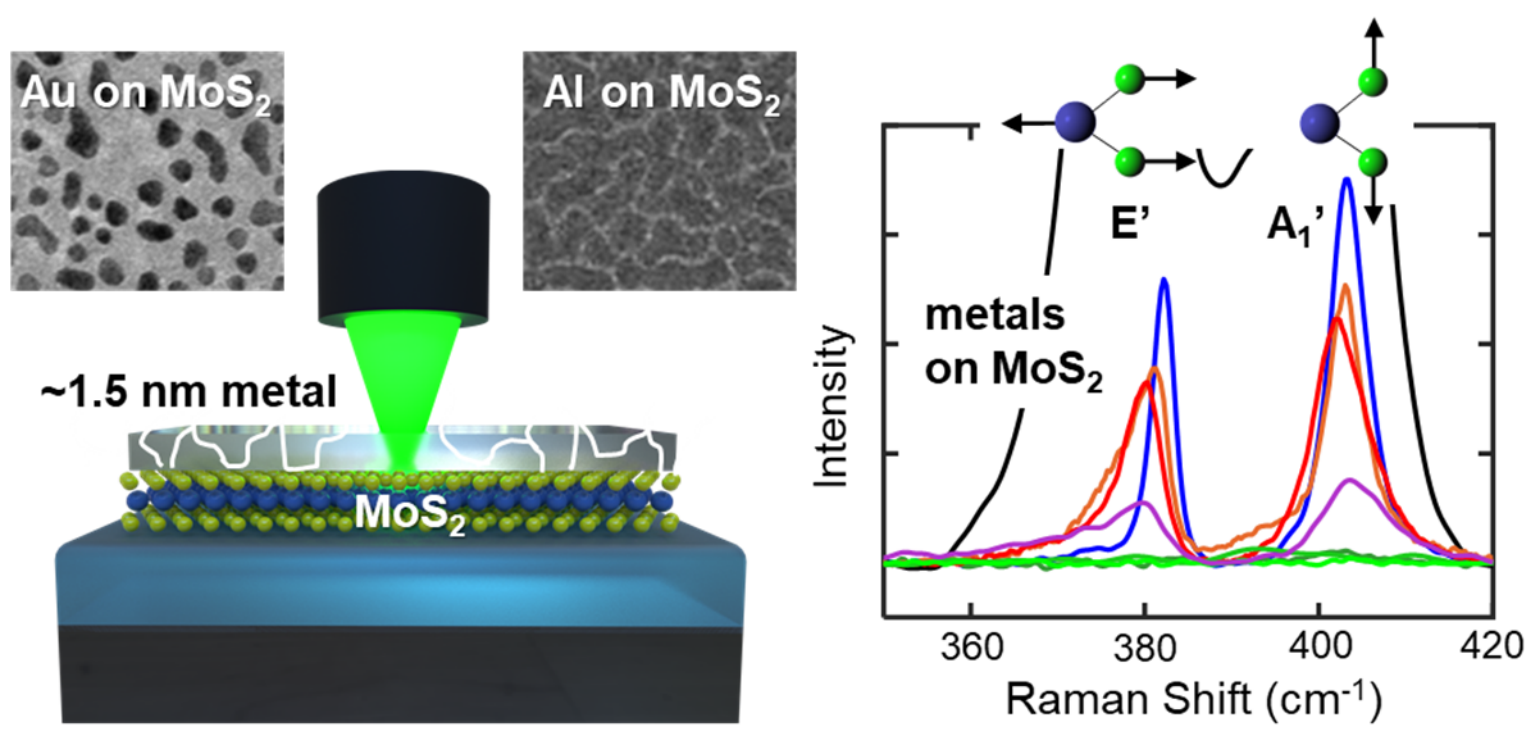

## REFERENCES

1. de Boer, B.; Hadipour, A.; Mandoc, M. M.; van Woudenbergh, T.; Blom, P. W. M., Tuning of metal work functions with self-assembled monolayers. *Adv. Mater.* **2005,** *17*, 621-625.
2. Lamport, Z. A.; Barth, K. J.; Lee, H.; Gann, E.; Engmann, S.; Chen, H.; Guthold, M.; McCulloch, I.; Anthony, J. E.; Richter, L. J.; DeLongchamp, D. M.; Jurchescu, O. D., A simple and robust approach to reducing contact resistance in organic transistors. *Nat. Commun.* **2018,** *9*, 5130.
3. Klauk, H., Will We See Gigahertz Organic Transistors? *Adv. Electron. Mater.* **2018,** *4*, 1700474.
4. Dimoulas, A.; Tsipas, P.; Sotiropoulos, A.; Evangelou, E. K., Fermi-level pinning and charge neutrality level in germanium. *Appl. Phys. Lett.* **2006,** *89*, 252110.
5. Himpsel, F. J.; Hollinger, G.; Pollak, R. A., Determination of the Fermi-Level Pinning Position at Si(111) Surfaces. *Phys. Rev. B* **1983,** *28*, 7014-7018.
6. Thompson, S. E.; Parthasarathy, S., Moore's law: the future of Si microelectronics. *Mater. Today* **2006,** *9*, 20-25.
7. English, C. D.; Shine, G.; Dorgan, V. E.; Saraswat, K. C.; Pop, E., Improved Contacts to $MoS_2$ Transistors by Ultra-High Vacuum Metal Deposition. *Nano Lett.* **2017,** *16*, 3824-3830.
8. Gong, C.; Colombo, L.; Wallace, R. M.; Cho, K., The Unusual Mechanism of Partial Fermi Level Pinning at Metal-$MoS_2$ Interfaces. *Nano Lett.* **2014,** *14*, 1714-1720.
9. Allain, A.; Kang, J.; Banerjee, K.; Kis, A., Electrical contacts to two-dimensional semiconductors. *Nat. Mater.* **2015,** *14*, 1195-1205.
10. Liu, Y.; Guo, J.; Zhu, E. B.; Liao, L.; Lee, S. J.; Ding, M. N.; Shakir, I.; Gambin, V.; Huang, Y.; Duan, X. F., Approaching the Schottky-Mott limit in van der Waals metal-semiconductor junctions. *Nature* **2018,** *557*, 696-700.

11. Mleczko, M. J.; Yu, A. C.; Smyth, C. M.; Chen, V.; Shin, Y. C.; Chatterjee, S.; Tsai, Y. C.; Nishi, Y.; Wallace, R. M.; Pop, E., Contact Engineering High-Performance n-Type $MoTe_2$ Transistors. *Nano Lett.* **2019,** *19*, 6352-6362.
12. English, C. D.; Smithe, K. K. H.; Xu, R. J.; Pop, E., Approaching Ballistic Transport in Monolayer $MoS_2$ Transistors with Self-Aligned 10 nm Top Gates. *IEEE Int. Electron Devices Meet.*, San Fransisco, CA, 2016.
13. Smithe, K. K. H.; English, C. D.; Suryavanshi, S. V.; Pop, E., High-Field Transport and Velocity Saturation in Synthetic Monolayer $MoS_2$. *Nano Lett.* **2018,** *18*, 4516-4522.
14. McClellan, C.; Suryavanshi, S.; English, C.; Smithe, K.; Bailey, C.; Grady, R.; Pop, E. 2D Semiconductor Transistor Device Trends. http://2d.stanford.edu/2D_Trends.
15. Lince, J. R.; Carre, D. J.; Fleischauer, P. D., Schottky-Barrier Formation on a Covalent Semiconductor without Fermi-Level Pinning: The Metal-$MoS_2$(0001) Interface. *Phys. Rev. B* **1987,** *36*, 1647-1656.
16. McDonnell, S.; Smyth, C.; Hinkle, C. L.; Wallace, R. M., $MoS_2$-Titanium Contact Interface Reactions. *ACS Appl. Mater. Interfaces* **2016,** *8*, 8289-8294.
17. McGovern, I. T.; Dietz, E.; Rotermund, H. H.; Bradshaw, A. M.; Braun, W.; Radlik, W.; Mcgilp, J. F., Soft-X-Ray Photoemission Spectroscopy of Metal Molybdenum-Disulfide Interfaces. *Surf. Sci.* **1985,** *152*, 1203-1212.
18. Smyth, C. M.; Addou, R.; McDonnell, S.; Hinkle, C. L.; Wallace, R. M., Contact Metal-$MoS_2$ Interfacial Reactions and Potential Implications on $MoS_2$-Based Device Performance. *J. Phys. Chem. C* **2016,** *120*, 14719-14729.
19. Wu, R. J.; Udyavara, S.; Ma, R.; Wang, Y.; Chhowalla, M.; Birol, T.; Koester, S. J.; Neurock, M.; Mkhoyan, K. A., Visualizing the metal-$MoS_2$ contacts in two-dimensional field-effect transistors with atomic resolution. *Phys. Rev. Mater.* **2019,** *3*, 111001.
20. Cooley, K. A.; Alsaadi, R.; Gurunathan, R. L.; Domask, A. C.; Kerstetter, L.; Saidi, W. A.; Mohney, S. E., Room-temperature epitaxy of metal thin films on tungsten diselenide. *J. Cryst. Growth* **2019,** *505*, 44-51.
21. Domask, A. C.; Cooley, K. A.; Kabius, B.; Abraham, M.; Mohney, S. E., Room Temperature van der Waals Epitaxy of Metal Thin Films on Molybdenum Disulfide. *Cryst. Growth Des.* **2018,** *18*, 3494-3501.
22. Golasa, K.; Grzeszczyk, M.; Binder, J.; Bozek, R.; Wysmolek, A.; Babinski, A., The disorder-induced Raman scattering in Au/$MoS_2$ heterostructures. *AIP Adv.* **2015,** *5*, 077120.
23. Gong, C.; Huang, C. M.; Miller, J.; Cheng, L. X.; Hao, Y. F.; Cobden, D.; Kim, J.; Ruoff, R. S.; Wallace, R. M.; Cho, K.; Xu, X. D.; Chabal, Y. J., Metal Contacts on Physical Vapor Deposited Monolayer $MoS_2$. *ACS Nano* **2013,** *7*, 11350-11357.
24. Moe, Y. A.; Sun, Y. H.; Ye, H. Y.; Liu, K.; Wang, R. M., Probing Evolution of Local Strain at $MoS_2$-Metal Boundaries by Surface-Enhanced Raman Scattering. *ACS Appl. Mater. Interfaces* **2018,** *10*, 40246-40254.
25. Sun, Y. H.; Liu, K.; Hong, X. P.; Chen, M.; Kim, J.; Shi, S. F.; Wu, J. Q.; Zettl, A.; Wang, F., Probing Local Strain at $MX_2$-Metal Boundaries with Surface Plasmon-Enhanced Raman Scattering. *Nano Lett.* **2014,** *14*, 5329-5334.
26. Chakraborty, B.; Bera, A.; Muthu, D. V. S.; Bhowmick, S.; Waghmare, U. V.; Sood, A. K., Symmetry-dependent phonon renormalization in monolayer $MoS_2$ transistor. *Phys. Rev. B* **2012,** *85*, 161403.
27. Carvalho, B. R.; Wang, Y. X.; Mignuzzi, S.; Roy, D.; Terrones, M.; Fantini, C.; Crespi, V. H.; Malard, L. M.; Pimenta, M. A., Intervalley scattering by acoustic phonons in two-dimensional $MoS_2$ revealed by double-resonance Raman spectroscopy. *Nat. Commun.* **2017,** *8*, 14670.
28. Frey, G. L.; Tenne, R.; Matthews, M. J.; Dresselhaus, M. S.; Dresselhaus, G., Raman and resonance Raman investigation of $MoS_2$ nanoparticles. *Phys. Rev. B* **1999,** *60*, 2883-2892.
29. Rice, C.; Young, R. J.; Zan, R.; Bangert, U.; Wolverson, D.; Georgiou, T.; Jalil, R.; Novoselov, K. S., Raman-scattering measurements and first-principles calculations of strain-induced phonon shifts in monolayer $MoS_2$. *Phys. Rev. B* **2013,** *87*, 081307.

30. Ataca, C.; Topsakal, M.; Aktürk, E.; Ciraci, S., A Comparative Study of Lattice Dynamics of Three- and Two-Dimensional MoS2. *J. Phys. Chem. C* **2011,** *115*, 16354-16361.
31. Michaelson, H. B., Work Function of Elements and Its Periodicity. *J. Appl. Phys.* **1977,** *48*, 4729-4733.
32. Uda, M.; Nakamura, A.; Yamamoto, T.; Fujimoto, Y., Work function of polycrystalline Ag, Au and Al. *J. Electron Spectrosc. Relat. Phenom.* **1998,** *88*, 643-648.
33. Smithe, K. K. H.; English, C. D.; Suryavanshi, S. V.; Pop, E., Intrinsic electrical transport and performance projections of synthetic monolayer $MoS_2$ devices. *2D Mater.* **2017,** *4*, 011009.
34. Smithe, K. K. H.; Suryavanshi, S. V.; Muñoz Rojo, M.; Tedjarati, A. D.; Pop, E., Low variability in synthetic monolayer $MoS_2$ devices. *ACS Nano* **2017,** *11*, 8456-8463.
35. Zhong, H.; Quhe, R.; Wang, Y.; Ni, Z.; Ye, M.; Song, Z.; Pan, Y.; Yang, J.; Yang, L.; Lei, M., Interfacial properties of monolayer and bilayer $MoS_2$ contacts with metals: beyond the energy band calculations. *Sci. Rep.* **2016,** *6*, 21786.
36. Lee, H.; Deshmukh, S.; Wen, J.; Costa, V. Z.; Schuder, J. S.; Sanchez, M.; Ichimura, A. S.; Pop, E.; Wang, B.; Newaz, A. K. M., Layer-Dependent Interfacial Transport and Optoelectrical Properties of $MoS_2$ on Ultraflat Metals. *ACS Appl. Mater. Interfaces* **2019,** *11*, 31543-31550.
37. Ohguro, T.; Nakamura, S.; Saito, M.; Ono, M.; Harakawa, H.; Morifuji, E.; Yoshitomi, T.; Morimoto, T.; Momose, H. S.; Katsumata, Y.; Iwai, H., Ultra-shallow junction and salicide techniques for advanced CMOS devices. *Elec Soc S* **1997,** *1997*, 275-295.
38. Freedy, K. M.; Giri, A.; Foley, B. M.; Barone, M. R.; Hopkins, P. E.; McDonnell, S., Titanium contacts to graphene: process-induced variability in electronic and thermal transport. *Nanotechnol.* **2018,** *29*, 145201.
39. Domask, A. C.; Gurunathan, R. L.; Mohney, S. E., Transition Metal-$MoS_2$ Reactions: Review and Thermodynamic Predictions. *J. Electron. Mater.* **2015,** *44*, 4065-4079.
40. Michail, A.; Delikoukos, N.; Parthenios, J.; Galiotis, C.; Papagelis, K., Optical detection of strain and doping inhomogeneities in single layer $MoS_2$. *Appl. Phys. Lett.* **2016,** *108*, 173102.
41. Kim, S. Y.; Yang, H. I.; Choi, W., Photoluminescence quenching in monolayer transition metal dichalcogenides by $Al_2O_3$ encapsulation. *Appl. Phys. Lett.* **2018,** *113*, 133104.
42. Leonhardt, A.; Chiappe, D.; Afanas'ev, V. V.; El Kazzi, S.; Shlyakhov, I.; Conard, T.; Franquet, A.; Huyghebaert, C.; de Gendt, S., Material-Selective Doping of 2D TMDC through $Al_xO_y$ Encapsulation. *ACS Appl. Mater. Interfaces* **2019,** *11*, 42697-42707.
43. McClellan, C. J.; Yalon, E.; Smithe, K. K. H.; Suryavanshi, S. V.; Pop, E., Effective n-type Doping of Monolayer $MoS_2$ by $AlO_x$. *IEEE Device Res. Conf.* **2017**.
44. Nickel, N. H.; Lengsfeld, P.; Sieber, I., Raman spectroscopy of heavily doped polycrystalline silicon thin films. *Phys. Rev. B* **2000,** *61*, 15558-15561.
45. Cerdeira, F.; Fjeldly, T. A.; Cardona, M., Effect of Free Carriers on Zone-Center Vibrational Modes in Heavily Doped P-Type Si .2. Optical Modes. *Phys. Rev. B* **1973,** *8*, 4734-4745.
46. Cerdeira, F.; Cardona, M., Effect of Carrier Concentration on Raman Frequencies of Si and Ge. *Phys. Rev. B* **1972,** *5*, 1440-1454.
47. Rao, A. M.; Eklund, P. C.; Bandow, S.; Thess, A.; Smalley, R. E., Evidence for charge transfer in doped carbon nanotube bundles from Raman scattering. *Nature* **1997,** *388*, 257-259.
48. Mignuzzi, S.; Pollard, A. J.; Bonini, N.; Brennan, B.; Gilmore, I. S.; Pimenta, M. A.; Richards, D.; Roy, D., Effect of disorder on Raman scattering of single-layer $MoS_2$. *Phys. Rev. B* **2015,** *91*, 195411.
49. Shi, W.; Zhang, X.; Li, X. L.; Qiao, X. F.; Wu, J. B.; Zhang, J.; Tan, P. H., Phonon Confinement Effect in Two-dimensional Nanocrystallites of Monolayer $MoS_2$ to Probe Phonon Dispersion Trends Away from Brillouin-Zone Center. *Chinese Phys. Lett.* **2016,** *33*, 057801.
50. He, Z.; Zhao, R.; Chen, X.; Chen, H.; Zhu, Y.; Su, H.; Huang, S.; Xue, J.; Dai, J.; Cheng, S.; Liu, M.; Wang, X.; Chen, Y., Defect Engineering in Single-Layer $MoS_2$ Using Heavy Ion Irradiation. *ACS Appl. Mater. Interfaces* **2018,** *10*, 42524-42533.

51. Chen, Y.; Huang, S. X.; Ji, X.; Adepalli, K.; Yin, K. D.; Ling, X.; Wang, X. W.; Xue, J. M.; Dresselhaus, M.; Kong, J.; Yildiz, B., Tuning Electronic Structure of Single Layer $MoS_2$ through Defect and Interface Engineering. *ACS Nano* **2018,** *12*, 2569-2579.
52. Parkin, W. M.; Balan, A.; Liang, L. B.; Das, P. M.; Lamparski, M.; Naylor, C. H.; Rodriguez-Manzo, J. A.; Johnson, A. T. C.; Meunier, V.; Drndic, M., Raman Shifts in Electron-Irradiated Monolayer $MoS_2$. *ACS Nano* **2016,** *10*, 4134-4142.
53. Schroeder, D. K., *Semiconductor Material and Device Characterization*; 3rd Edition. John Wiley & Sons: New Jersey, **2006**.
54. English, C. D.; Shine, G.; Dorgan, V. E.; Saraswat, K. C.; Pop, E., Improved Contacts to $MoS_2$ Transistors by Ultra-High Vacuum Metal Deposition. *Nano Lett.* **2016,** *16*, 3824-3830.
55. Bampoulis, P.; van Bremen, R.; Yao, Q. R.; Poelsema, B.; Zandvliet, H. J. W.; Sotthewes, K., Defect Dominated Charge Transport and Fermi Level Pinning in $MoS_2$/Metal Contacts. *ACS Appl. Mater. Interfaces* **2017,** *9*, 19278-19286.
56. Cheng, Z. H.; Cardenas, J. A.; McGuire, F.; Najmaei, S.; Franklin, A. D., Modifying the Ni-$MoS_2$ Contact Interface Using a Broad-Beam Ion Source. *IEEE Electron Device Lett.* **2016,** *37*, 1234-1237.
57. Passi, V.; Gahoi, A.; Marin, E. G.; Cusati, T.; Fortunelli, A.; Iannaccone, G.; Fiori, G.; Lemme, M. C., Ultralow Specific Contact Resistivity in Metal-Graphene Junctions via Contact Engineering. *Adv. Mater. Interfaces* **2019,** *6*, 1801285.
58. Rai, A.; Valsaraj, A.; Movva, H. C. P.; Roy, A.; Ghosh, R.; Sonde, S.; Kang, S. W.; Chang, J. W.; Trivedi, T.; Dey, R.; Guchhait, S.; Larentis, S.; Register, L. F.; Tutuc, E.; Banerjee, S. K., Air Stable Doping and Intrinsic Mobility Enhancement in Monolayer Molybdenum Disulfide by Amorphous Titanium Suboxide Encapsulation. *Nano Lett.* **2015,** *15*, 4329-4336.
59. Liu, W.; Kang, J.; Sarkar, D.; Khatami, Y.; Jena, D.; Banerjee, K., Role of metal contacts in designing high-performance monolayer n-type $WSe_2$ field effect transistors. *Nano Lett.* **2013,** *13*, 1983-1990.
60. Conley, H. J.; Wang, B.; Ziegler, J. I.; Haglund, R. F., Jr.; Pantelides, S. T.; Bolotin, K. I., Bandgap engineering of strained monolayer and bilayer $MoS_2$. *Nano Lett.* **2013,** *13*, 3626-3630.
61. Dai, Z.; Liu, L.; Zhang, Z., Strain Engineering of 2D Materials: Issues and Opportunities at the Interface. *Adv. Mater.* **2019,** *31*, e1805417.
62. Li, H.; Contryman, A. W.; Qian, X. F.; Ardakani, S. M.; Gong, Y. J.; Wang, X. L.; Weisse, J. M.; Lee, C. H.; Zhao, J. H.; Ajayan, P. M.; Li, J.; Manoharan, H. C.; Zheng, X. L., Optoelectronic crystal of artificial atoms in strain-textured molybdenum disulphide. *Nat. Commun.* **2015,** *6*, 7381.
63. Zhu, C. R.; Wang, G.; Liu, B. L.; Marie, X.; Qiao, X. F.; Zhang, X.; Wu, X. X.; Fan, H.; Tan, P. H.; Amand, T.; Urbaszek, B., Strain tuning of optical emission energy and polarization in monolayer and bilayer $MoS_2$. *Phys. Rev. B* **2013,** *88*, 121301.
64. Amani, M.; Chin, M. L.; Mazzoni, A. L.; Burke, R. A.; Najmaei, S.; Ajayan, P. M.; Lou, J.; Dubey, M., Growth-substrate induced performance degradation in chemically synthesized monolayer $MoS_2$ field effect transistors. *Appl. Phys. Lett.* **2014,** *104*, 203506.
65. Ahn, G. H.; Amani, M.; Rasool, H.; Lien, D. H.; Mastandrea, J. P.; Ager, J. W.; Dubey, M.; Chrzan, D. C.; Minor, A. M.; Javey, A., Strain-engineered growth of two-dimensional materials. *Nat. Commun.* **2017,** *8*, 608.
66. Hosseini, M.; Elahi, M.; Pourfath, M.; Esseni, D., Very large strain gauges based on single layer $MoSe_2$ and $WSe_2$ for sensing applications. *Appl. Phys. Lett.* **2015,** *107*, 253503.
67. Lee, B.; Park, J.; Han, G. H.; Ee, H. S.; Naylor, C. H.; Liu, W. J.; Johnson, A. T. C.; Agarwal, R., Fano Resonance and Spectrally Modified Photoluminescence Enhancement in Monolayer $MoS_2$ Integrated with Plasmonic Nanoantenna Array. *Nano Lett.* **2015,** *15*, 3646-3653.
68. McGilp, J. F., On Predicting the Chemical-Reactivity of Metal-Semiconductor Interfaces. *J. Phys. C: Solid State Phys.* **1984,** *17*, 2249-2254.
69. Komsa, H. P.; Kotakoski, J.; Kurasch, S.; Lehtinen, O.; Kaiser, U.; Krasheninnikov, A. V., Two-Dimensional Transition Metal Dichalcogenides under Electron Irradiation: Defect Production and Doping. *Phys. Rev. Lett.* **2012,** *109*.

70. Saitoh, M.; Nakabayashi, Y.; Uchida, K.; Numata, T., Short-Channel Performance Improvement by Raised Source/Drain Extensions With Thin Spacers in Trigate Silicon Nanowire MOSFETs. *IEEE Electron Device Lett.* **2011,** *32*, 273-275.
71. Yang, Y. R.; Breil, N.; Yang, C. Y.; Hsieh, J.; Chiang, F.; Colombeau, B.; Guo, B. N.; Shim, K. H.; Variam, N.; Leung, G.; Hebb, J.; Sharma, S.; Ni, C. N.; Ren, J.; Wen, J.; Park, J. H.; Chen, H.; Chen, S.; Hou, M.; Tsai, D., et al. Ultra low p-type SiGe contact resistance FinFETs with Ti silicide liner using cryogenic contact implantation amorphization and Solid-Phase Epitaxial Regrowth (SPER), *IEEE Symp. VLSI Technol.*, Honolulu, HI, Honolulu, HI, **2016**; pp 74-75.

# Supporting Information

# Uncovering the Effects of Metal Contacts on Monolayer $MoS_2$

**Kirstin Schauble,[1] Dante Zakhidov,[2] Eilam Yalon,[1,3] Sanchit Deshmukh,[1] Ryan W. Grady,[1] Kayla A. Cooley,[4] Connor J. McClellan,[1] Sam Vaziri,[1] Donata Passarello,[5] Suzanne E. Mohney,[4] Michael F. Toney,[5] A. K. Sood,[6] Alberto Salleo,[2] and Eric Pop[1,2,7,*]**

[1]Department of Electrical Engineering, Stanford Univ., Stanford, CA 94305, USA. [*]epop@stanford.edu
[2]Department of Materials Science and Engineering, Stanford Univ., Stanford, CA 94305, USA
[3]Present address: Andrew and Erna Viterbi Department of Electrical Engineering, Technion 3200003, Israel
[4]Department of Materials Science and Engineering, The Pennsylvania State Univ., University Park, PA 16802, USA
[5]Stanford Synchrotron Radiation Lightsource, SLAC National Accelerator Laboratory, Menlo Park, USA
[6]Department of Physics, India Institute of Science, Bangalore 560012, India.
[7]Precourt Institute for Energy, Stanford Univ., Stanford, CA 94305, USA

## 1. Raman Analysis Details

For Raman analysis (data shown in main text Figure 1b), we fit the $MoS_2$ $A_1$' peaks with a Lorentzian. The $MoS_2$ E' peaks are fit as follows: for bare $MoS_2$, a Lorentzian is fit. For monolayer $MoS_2$ covered by Ti, Sc, and Y, the $MoS_2$ peaks disappear (Figure S1). For $MoS_2$ covered by ultrathin Al and Ni, an asymmetric Fano line shape is fit due to observed doping and damage effects (Figure S2). For $MoS_2$ covered by Au and Ag, two separate Lorentzian peaks are fit because these metals grow in islands on $MoS_2$ (see Figures S2d and S5), where one peak is from the $MoS_2$ not contacted by metal, and the other red-shifted peak is from the $MoS_2$ directly contacted by metal. For these metals, the position and full width at half maximum (FWHM) of the *shifted* peak ($MoS_2$ contacted by metal) are listed in Table 1 and Figure 4c of the main text.

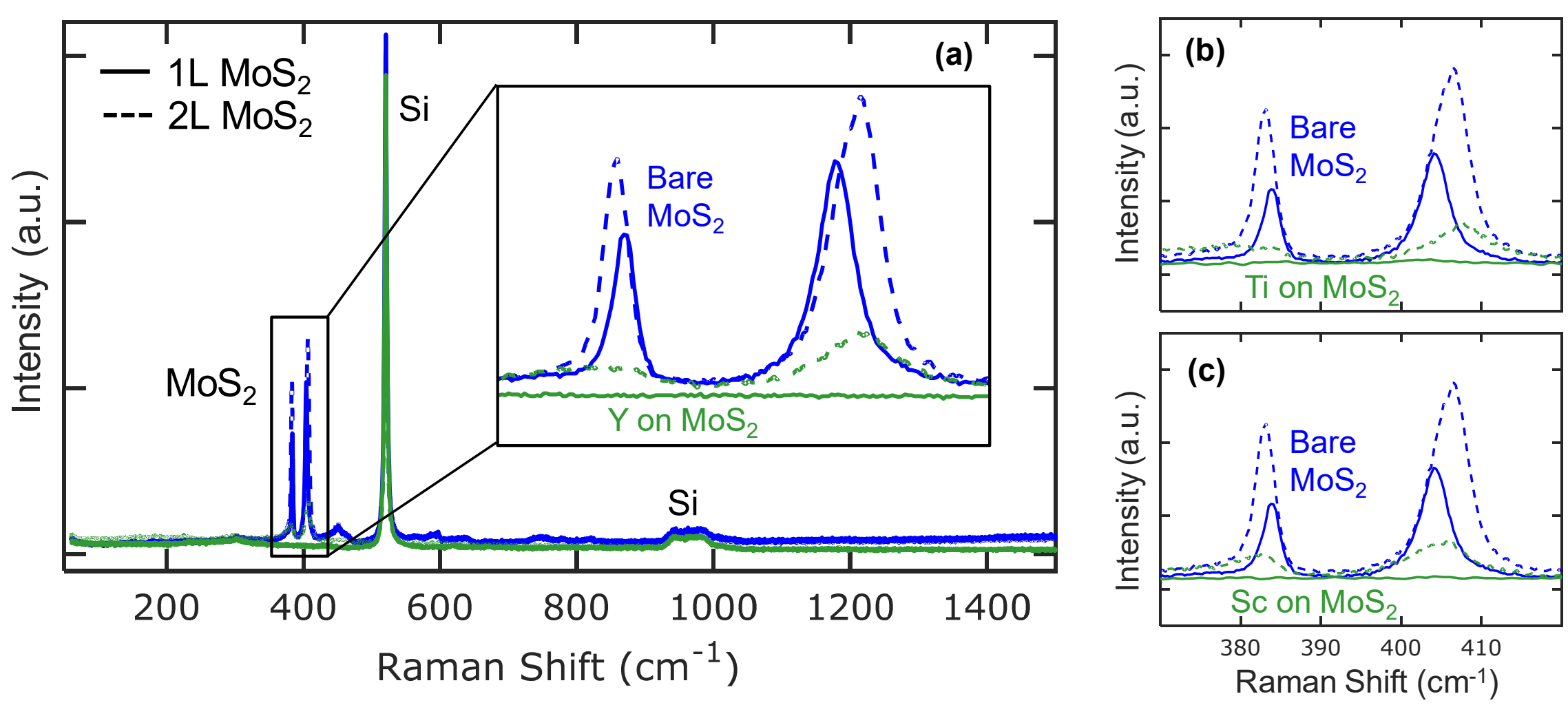

**Figure S1:** (a) Raman spectra of monolayer (1L, solid lines) and bilayer (2L, dashed lines) $MoS_2$ bare (blue) and coated with 1.5 nm Y (green), on our $SiO_2$/Si substrates. The inset shows zoomed in regions of the E' and $A_1$' $MoS_2$ peaks ($E_g$ and $A_{1g}$ in bilayer, respectively), where the Raman signal is present for Y on 2L $MoS_2$ but not 1L, suggesting Y reacts with the top layer but not fully with the bottom layer of 2L $MoS_2$. Si peaks remain present after Y deposition, which rules out laser reflection from the metal as the cause of the $MoS_2$ signal disappearance. These data are representative of ~1.5 nm (b) Ti and (c) Sc on $MoS_2$.

The Fano profile in Figure S2 is defined by $\alpha(\omega) = \alpha_0(q + x)^2/(1 + x^2)$ where $\alpha_0$ is the prefactor, $x = 2(\omega - \omega_P)/\gamma$, $\omega$ is frequency, $\omega_P$ is bare phonon frequency, $\gamma$ is linewidth, and $q$ is the symmetry parameter which depends on electron-phonon coupling strength.[1] A smaller absolute value $|q|$ means more asymmetry and heavier doping. Heavy *n*-type doping shifts the Fermi level near the conduction band, causing interactions between the continuum of electronic states and optical phonons, and thus asymmetry in the Raman peak.

The E' peak of bare $MoS_2$ has a large $|q|$ showing the most symmetry, as expected. Al-coated $MoS_2$ has a small $|q|$ and thus large asymmetry, implying heavy doping. In addition, because the mobility under the Al-coated samples remains unchanged but a large negative $V_T$ shift is observed vs. the bare samples (see main text Figure 3b), we can ascribe the observed line asymmetry to *n*-type doping and not $MoS_2$ damage.

Ni-coated $MoS_2$ has even lower $|q|$ than Al, which would imply heavier $MoS_2$ doping based on the Fano fit. However, Ni-coated transistors also show the lowest mobility (see main text Figure 4) which suggests $MoS_2$ sustained the most damage during the evaporation of ultrathin Ni. Because E' peak broadening and asymmetry can be caused by damage (leading to phonon confinement and contribution of dispersive TO modes)[2, 3] we conclude that the asymmetric E' peak of Ni-coated $MoS_2$ is most likely due to damage.

Finally, the E' Raman peak of Au-coated $MoS_2$ (Figure S2d) shows a poor Fano fit and is better fit by two separate Lorentzian peaks, as explained above. These two Lorentzians (dashed gray lines) roughly correspond to the LO and TO phonon branches (partly activated by disorder),[2, 3] although a quantitative analysis of their asymmetry and the defect density induced is left to be pursued in future work.

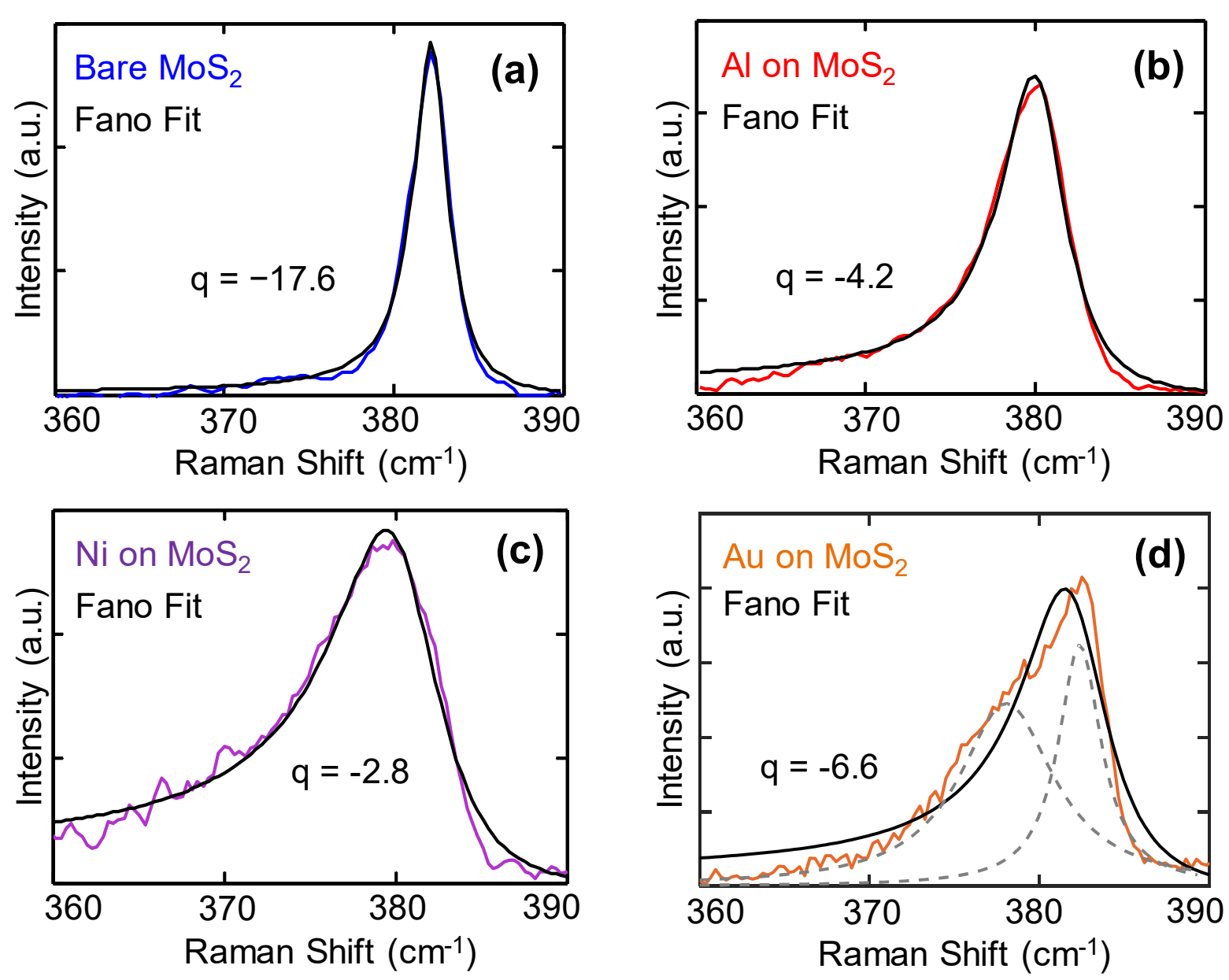

**Figure S2:** Fano line fits for the E' Raman peak of (a) bare $MoS_2$, (b) Al, (c) Ni, and (d) Au on $MoS_2$. The raw data are colored, the Fano fits are solid black lines, the Lorentzian fits in (d) are dashed gray lines.

## 2. XPS Analysis Details

All XPS spectra are normalized in binding energy (B.E.) by the $sp^3$ C 1s spectra at 284.8 eV, and normalized in height by the tallest peak in the spectral window. The bare $MoS_2$ XPS spectra (blue) are shown on the bottom of Figure S3, and the metal-coated $MoS_2$ spectra are listed in order of increasing metal work function. The spectra are color-coded based on the observed reactions, described below. We note that the XPS spot size (~200 μm) is much larger than the Raman spot size (<1 μm), so the sparse (~10%) bilayer overgrowth regions of $MoS_2$ inevitably contribute signal to these data (additional details in Methods).

First, the bare $MoS_2$ spectra (plotted in blue) are used as a reference, where the Mo 3d and S 2p spectra are indicative of $MoS_2$, and the O 1s spectrum shows an $SiO_2$ peak from the substrate. Second, Y, Sc, and Ti (plotted in green) react with $MoS_2$ and oxidize. For $MoS_2$ coated with these metals, a shift in the Mo 3d spectrum is seen, indicating sulfur-deficient $MoS_{2-x}$ and metallic Mo are left behind after the metals react with the sulfur from $MoS_2$ (detailed peak fitting of Ti on $MoS_2$ is provided in Figure 2 of the main text). These metals also induce broadening of the $MoS_2$ S 2p spectra, indicating metal-sulfide bonding. Note that the Y-coated $MoS_2$ S 2p spectrum is washed out by a large Y peak. We also see a shift in the O 1s spectra, which indicates signal from Ti, Sc, and Y oxides instead of the $SiO_2$ substrate. Third, Al and Ni (plotted in purple) oxidize but do not react with $MoS_2$. The Mo 3d and S 2p spectra remain unchanged, indicating that $MoS_2$ is still present. The O 1s spectra are shifted, indicating that the signal is from Al and Ni oxides instead of the $SiO_2$ substrate. Lastly, Ag and Au (plotted in black) do not react with $MoS_2$ or oxidize. The Mo 3d, S 2p, and O 1s spectra remain relatively unchanged from the bare $MoS_2$ spectrum. The S 2p spectrum of Ag-capped $MoS_2$ shows broadening, suggesting sulfurization of Ag. However, since the Mo 3d peak is unaffected, this is most likely due to Ag tarnishing from sulfur oxides in ambient air.

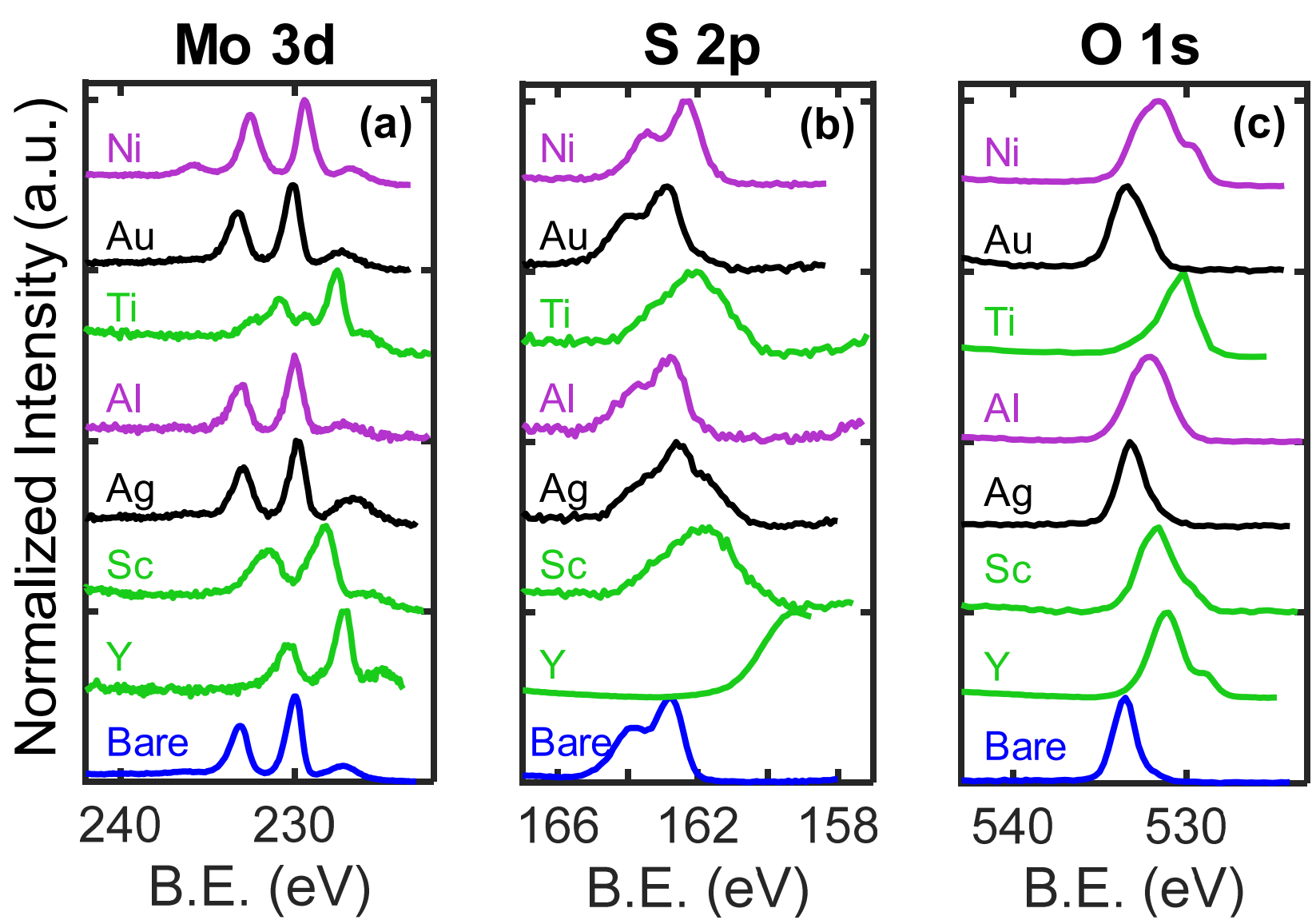

**Figure S3:** (a) XPS Mo 3d spectra shows doublet Mo 3d5/2 and Mo 3d3/2 peaks. (b) XPS S 2p spectra shows doublet S 2p3/2 and S 2p1/2 peaks. (c) XPS O 1s spectra shows $SiO_2$ or other oxidized metals.

## 3. SAED and TEM Data

Figure S4 shows selected area electron diffraction (SAED) patterns for $MoS_2$ coated with ultrathin Ag and Ni (see Methods for sample preparation on 20 nm $SiO_2$ TEM grids). For the Ag-coated sample, in addition to polycrystalline $MoS_2$ (yellow) and Ag (green), a secondary crystalline phase is present which can be attributed to $Ag_2S$ (purple) and potentially $Mo_8O_{23}$ (grey). For the Ni-coated sample, there is no evidence of crystalline phases besides $MoS_2$ indicating that Ni is oxidizing into an amorphous phase. These results are consistent with the XPS data in Figure S3.

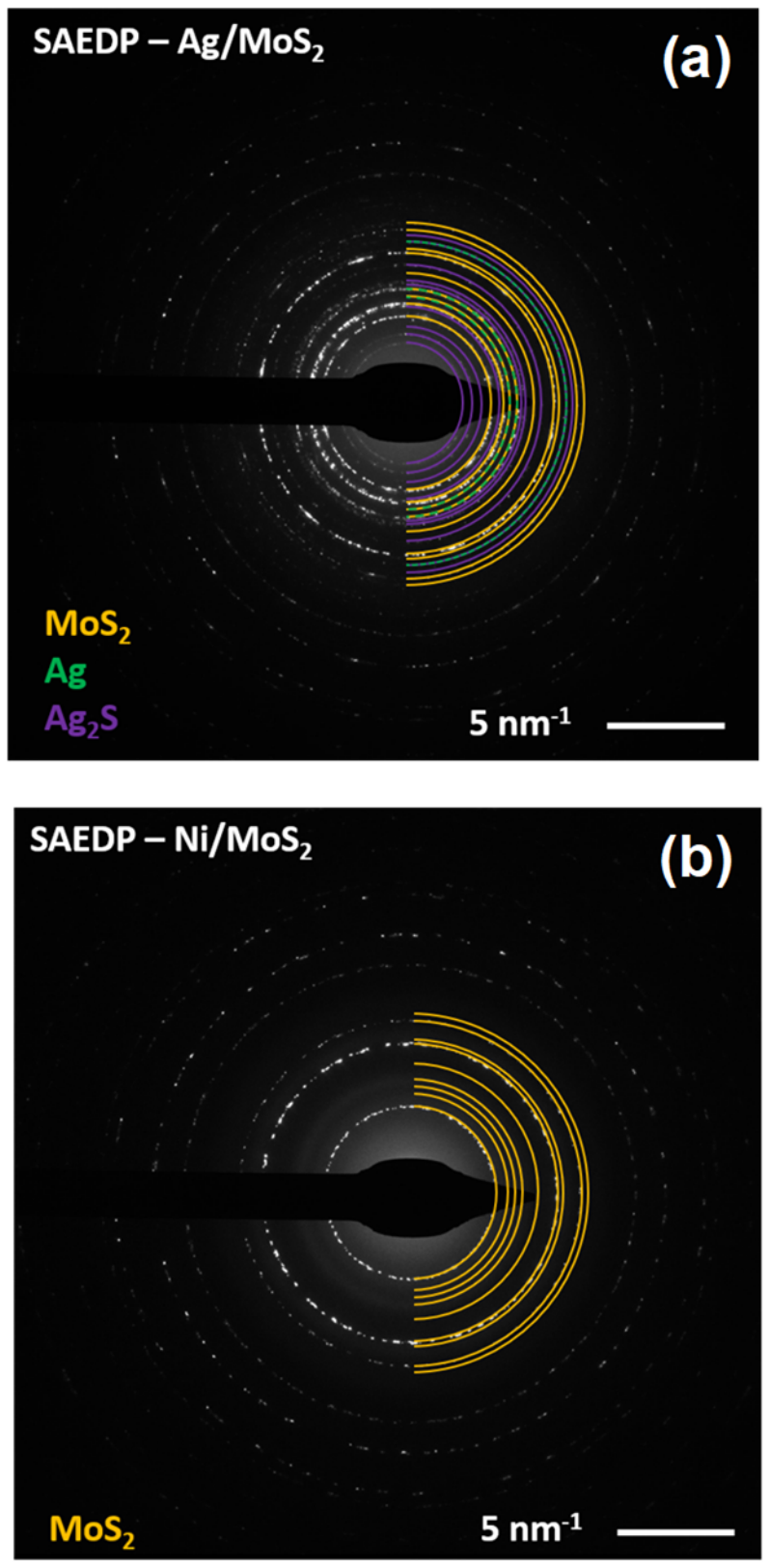

| Measured Ring Radius (nm$^{-1}$) | MoS$_2$ Planes | Ag Planes | Ag$_2$S Planes | Mo$_8$O$_{23}$ Planes |
|---|---|---|---|---|
| 2.50 | | | 100 | 020 |
| 2.87 | | | 110 | 302 |
| 3.23 | | | 002 | |
| 3.63 | 010 | | | 006 |
| 4.0 | | | 021 | |
| 4.13 | 012 | | | |
| 4.43 | 013 | 111 | | 60$\bar{2}$ |
| 4.77 | 006 | 002 | | |
| 4.93 | | | 022 | 504 |
| 5.10 | | | | 50$\bar{8}$ |
| 5.40 | 015 | | $\bar{2}$12 | |
| 5.77 | | | 220 | $\bar{5}$28 |
| 6.30 | 110 | | 12$\bar{3}$ | 800 |
| 6.53 | 112 | | | 820 |
| 6.80 | | 022 | 10$\bar{4}$ | |
| 7.07 | | | 004 | 00 12 |
| 7.30 | 018 | | $\bar{1}$41 | |
| 7.60 | 023 | | $\bar{3}$21 | 02 12 |

**Figure S4:** SAED pattern of (a) Ag-coated $MoS_2$ and (b) Ni-coated $MoS_2$. The table identifies SAED diffraction planes for each ring in the Ag/$MoS_2$ system. The interplanar spacing (not shown) for the given phases was determined from crystal structures on the Crystallography Open Database.[4] Ring radii are colored according to the SAED diffraction plane identification (using the calculated interplanar spacing values): yellow for $MoS_2$, green for Ag, and purple for $Ag_2S$. Rows with two colors indicate that diffraction planes exist with the given ring radius for both materials.

Figure S5 shows additional plan view TEMs, revealing Au and Ag non-uniformly "ball up" when deposited on $MoS_2$ into ~10 to 20 nm size islands. Au (also in main text Figure 4b) shows slightly denser nucleation on $SiO_2$ than $MoS_2$, whereas it is difficult to see contrast of $MoS_2$ edges at this scale on the Ag sample due to high contrast between $MoS_2$ and Ag. Al and Ni also deposit as discontinuous films on $MoS_2$, where Ni was found to completely oxidize but some unoxidized Al signal was present in EDS. Ag and Au did not oxidize, consistent with XPS (Figure S3). Al on $MoS_2$ (also in main text Figure 3c) has the largest island size, as well as increased continuity on $MoS_2$ than $SiO_2$. Ni islands are very small (< 10 nm) and nucleate with similar density on $MoS_2$ and $SiO_2$.

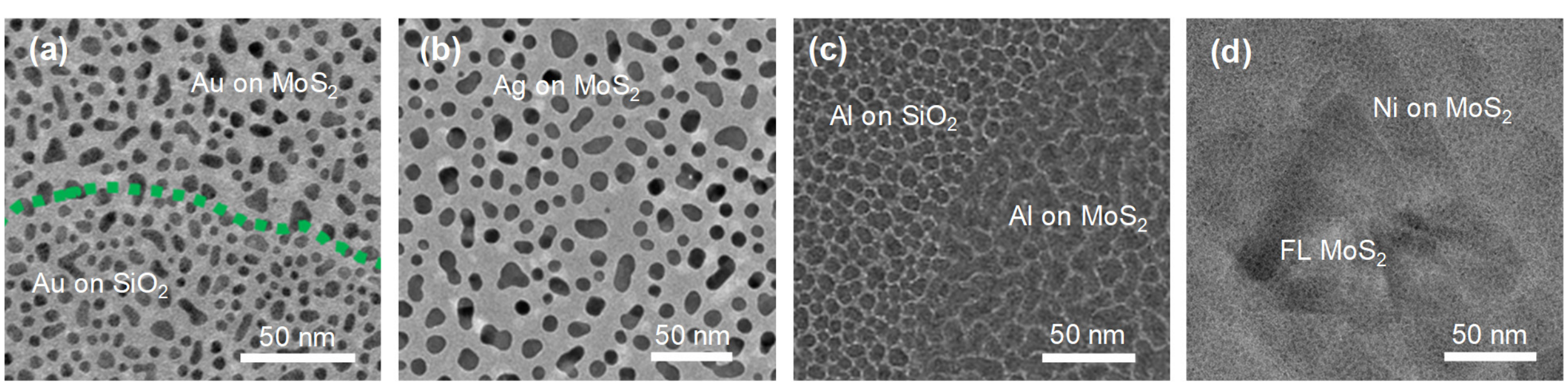

**Figure S5:** Plan view TEM of ultrathin (a) Au, (b) Ag, (c) Al, and (d) Ni on as-grown monolayer $MoS_2$.

## 4. Enthalpy of Formation Analysis

Table S1 displays enthalpy of formation of $MoS_2$, metal oxides, and metal sulfides, in units of kJ per mole of solid S for the sulfides and gaseous $O_2$ for the oxides (their most stable forms). Y, Sc, and Ti sulfides have enthalpies of formation that are more negative than $MoS_2$, meaning the sulfur atoms in $MoS_2$ prefer to bond with Y, Sc, and Ti rather than Mo, assuming small entropies of formation and no kinetic limitations. Ag, Au, and Ni have less negative enthalpies of formation with sulfur (or nonexistent in the case of Au), and thus their reactions with $MoS_2$ are thermodynamically unlikely (consistent with our XPS data).

Al has a similar work function to Ti and is thermodynamically expected to react with $MoS_2$, but none of our results indicate any reaction between Al and $MoS_2$. This has also been observed in previous studies,[5, 6] and can be explained by a larger kinetic barrier for reaction with sulfur, or by Al oxidation since $Al_2O_3$ has a much more negative enthalpy of formation (Table S1). In addition, it may be noteworthy that Al has the lowest melting temperature and requires the lowest power evaporation of all metals studied. Therefore, Al deposition may induce fewer $MoS_2$ defects than other metals during evaporation, which could prevent a reaction with $MoS_2$ (assuming that defects mediate the reaction).

We note that full thermodynamic analysis would involve the Gibbs free energy, which includes the entropy of formation, and the construction of a phase diagram.[7] However, many entropy values required for such an analysis are either unavailable or have large error bars, so we use enthalpy of formation for rapid but approximate insight. To make a fairer comparison, we report values per mole of solid S (sulfides) or gaseous $O_2$ (oxides), allowing the larger entropy term for gaseous $O_2$ to cancel in the comparison among oxides.

| | **Enthalpy of Formation (kJ/mol)** | | | | | |
|---|---|---|---|---|---|---|
| | S | Result | Ref. | O | Result | Ref. |
| Mo | **-138** | **$MoS_2$** | 8 | -452 | $MoO_3$ | 9 |
| Ni | -74 | $NiS_2$ | 8 | **-427** | **NiO** | 9 |
| Au | | -- | | | -- | |
| Ti | **-205** | **$TiS_2$** | 10 | **-889** | **$TiO_2$** | 9 |
| Al | -151 | AlS | 8 | **-1054** | **$Al_2O_3$** | 9 |
| Ag | **-32** | **$Ag_2S$** | 11 | -21 | $Ag_2O$ | 9 |
| Sc | **-774** | **$Sc_{0.8065}S$** | 12 | **-1088** | **$Sc_2O_3$** | 9 |
| Y | **-262** | **YS** | 13 | **-1121** | **$Y_2O_3$** | 9 |

**Table S1:** Thermodynamic enthalpy of formation for sulfides and oxides of the metals studied at room temperature. Bolded values are the reactions seen experimentally in XPS data (from Figure S3).

## 5. Additional Electrical Measurements

Figure S6 shows an example pseudo-TLM mobility extraction (following Smithe *et al.*[14]) for a bare $MoS_2$ channel, where the same analysis is done to calculate the mobility of the $MoS_2$ channel after depositing each non-reacting metal. The resulting mobility values for $MoS_2$ devices bare and coated with each non-reacting metal are reported in the main text Figure 4c, and are found to degrade drastically after deposition of Ni, Au, and Ag. Note that while the channel lengths here are too long to accurately extract contact resistance, mobility values can be estimated from the sheet resistance (slope of $R_{TOT}$ vs. $L$).

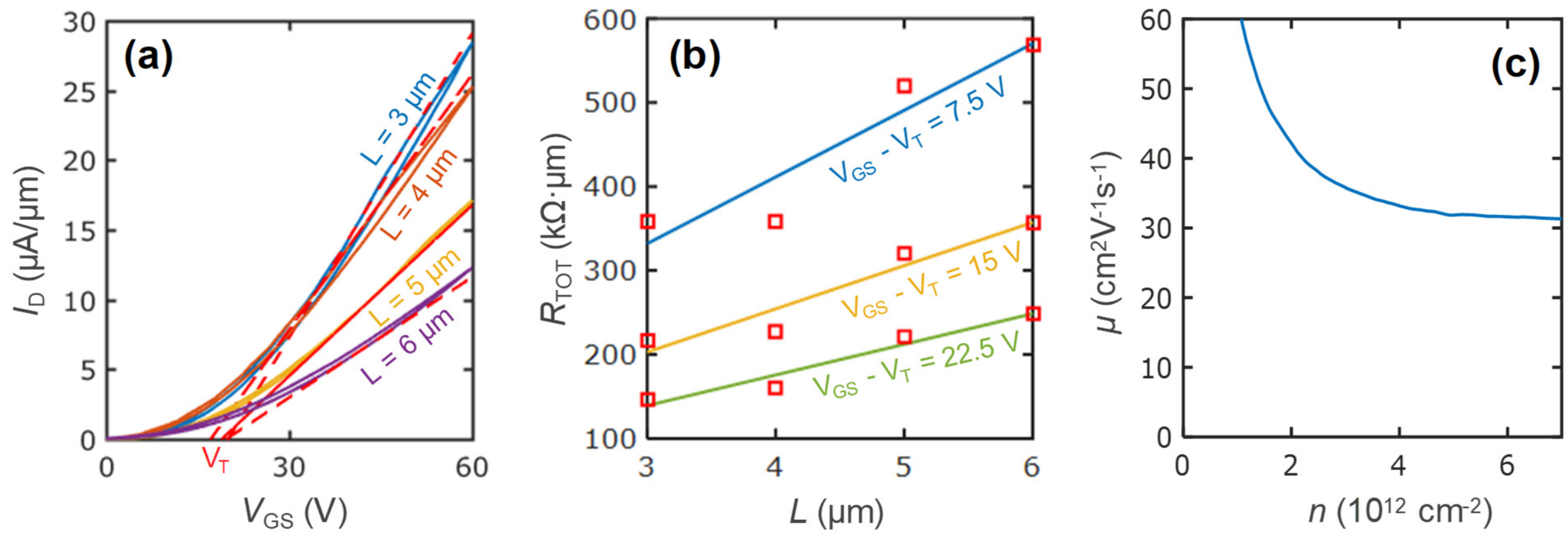

**Figure S6:** Example of pseudo-transfer length method (TLM)[14] extraction of mobility for bare (uncapped) monolayer $MoS_2$ devices. (a) Measured drain current vs. back-gate voltage ($I_D$ vs. $V_{GS}$) at $V_{DS}$ = 1 V for channel lengths 3 to 6 µm. Red dashed lines show linear extrapolation[14] to find threshold voltage $V_T$. (b) Total device resistance $R_{TOT}$ vs. channel length ($L$) measured by TLM at various gate overdrives ($V_{GS}$ - $V_T$). (c) Estimated effective mobility ($\mu$) vs. carrier density ($n$) based on $\mu = (qnR_{SH})^{-1}$, where sheet resistance $R_{SH}$ is the slope of $R_{TOT}$ vs. $L$ in (b). Figure 4c in the main text displays this mobility (for bare, Ag-, Au-, and Ni-coated samples) at $n \approx 7 \times 10^{12}$ cm$^{-2}$.

Figure S7 shows electrical measurements of $MoS_2$ capped with ultrathin Ti, Sc, and Y. These data are representative of several devices measured. Ti- and Y-coated $MoS_2$ devices lost nearly all gate dependence, and any conduction is due to byproducts of the reaction between the metal and $MoS_2$ (see Raman data in Figure S1 and XPS data in Figure S3). Sc-coated $MoS_2$ devices still show gate dependence, indicating some semiconducting behavior in the device channel, which must be due to either remaining $MoS_2$ (e.g. under bilayer regions) or byproducts of the Sc reaction with $MoS_2$ (ScS is an $n$-type conductor).[15] Additionally, we note the device channels had gone through several rounds of photolithography prior to ultrathin metal evaporation, so any residual photoresist may hinder interfacial reactions.

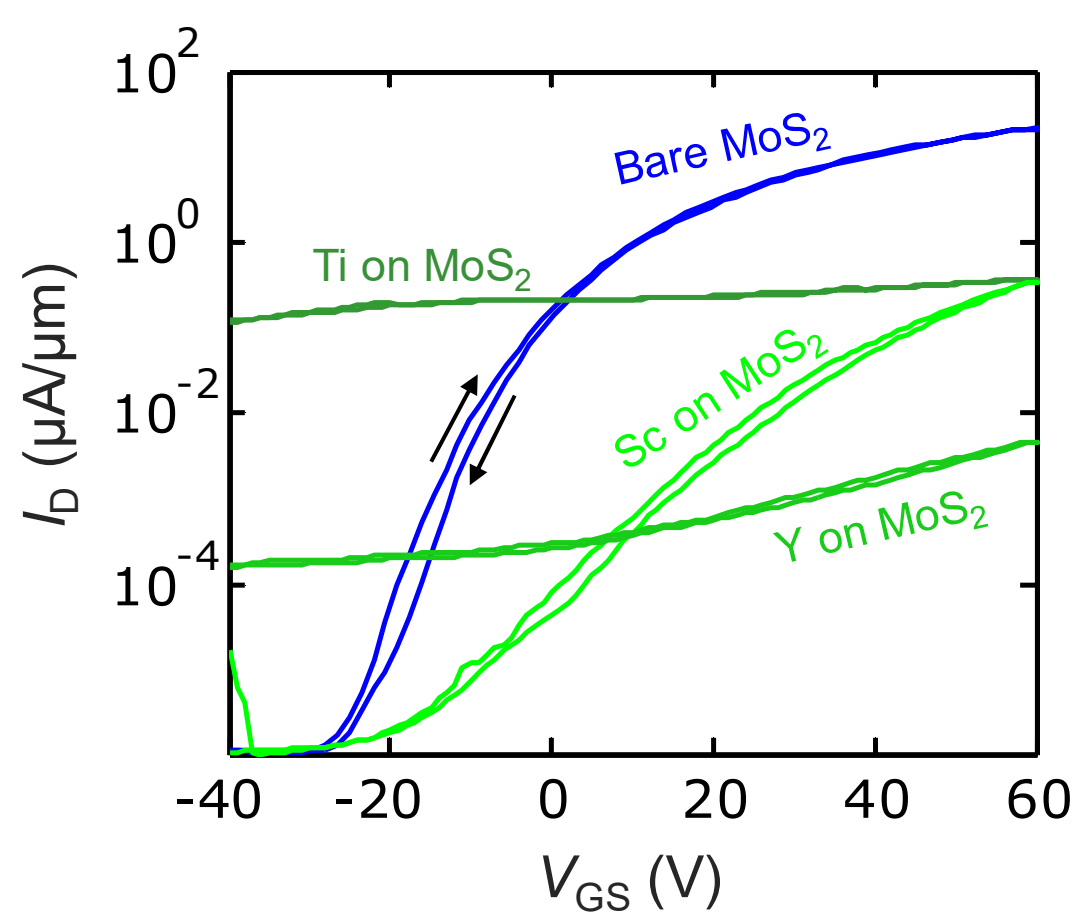

**Figure S7:** Measured drain current vs. back-gate voltage ($I_D$ vs. $V_{GS}$) of monolayer $MoS_2$ devices coated with ultrathin films (~1.5 nm) of low work function metals (Y, Sc, and Ti). $L$ = 3 µm, $V_{DS}$ = 1 V. Small arrows show the sweep direction, revealing repeatable measurements with minimal hysteresis. These data are representative of 5-10 devices measured for each kind of ultrathin metal coverage.

## 6. XRD Analysis Details

In order to calculate the built-in strain of our monolayer $MoS_2$ grown by chemical vapor deposition (CVD), we transfer (with the process described by Vaziri *et al.*[16]) the as-grown $MoS_2$ onto a fresh $SiO_2$/Si substrate and measure the strain before and after transferring (without any contact metals). This analysis assumes that transferring the $MoS_2$ releases its built-in strain. Figure S8 displays Raman and X-ray diffraction (XRD) data of as-grown and transferred $MoS_2$. Using Raman-based strain analysis, we find that the as-grown $MoS_2$ is tensile strained ~0.4 to 0.5%, based on a calibration of 4.5 $cm^{-1}$ E' peak shift per % biaxial strain as reported by Li *et al.*[17] XRD-based strain analysis uses Bragg's law to calculate the spacing between (10) rows of $MoS_2$ atoms in the 2D layer, $d = n\lambda/[2\sin(\theta)]$, where the X-ray wavelength $\lambda$ = 0.886 Å (14 keV) and $n$ is the order of reflection [here $n$ = 1 for the $MoS_2$ (10) rows, $n$ = 2 for $MoS_2$ (20) rows, etc.]. XRD analysis verifies that the Raman-based estimates of built-in $MoS_2$ tensile strain after CVD growth are accurate, as seen in Figure S8c. Raman and XRD analysis are in agreement for measuring built-in strain of bare $MoS_2$, however we show in the main text Figure 5 that Raman analysis is not predictive in determining strain of $MoS_2$ under contact metals, ostensibly due to the metal influence on the E' peak shift.

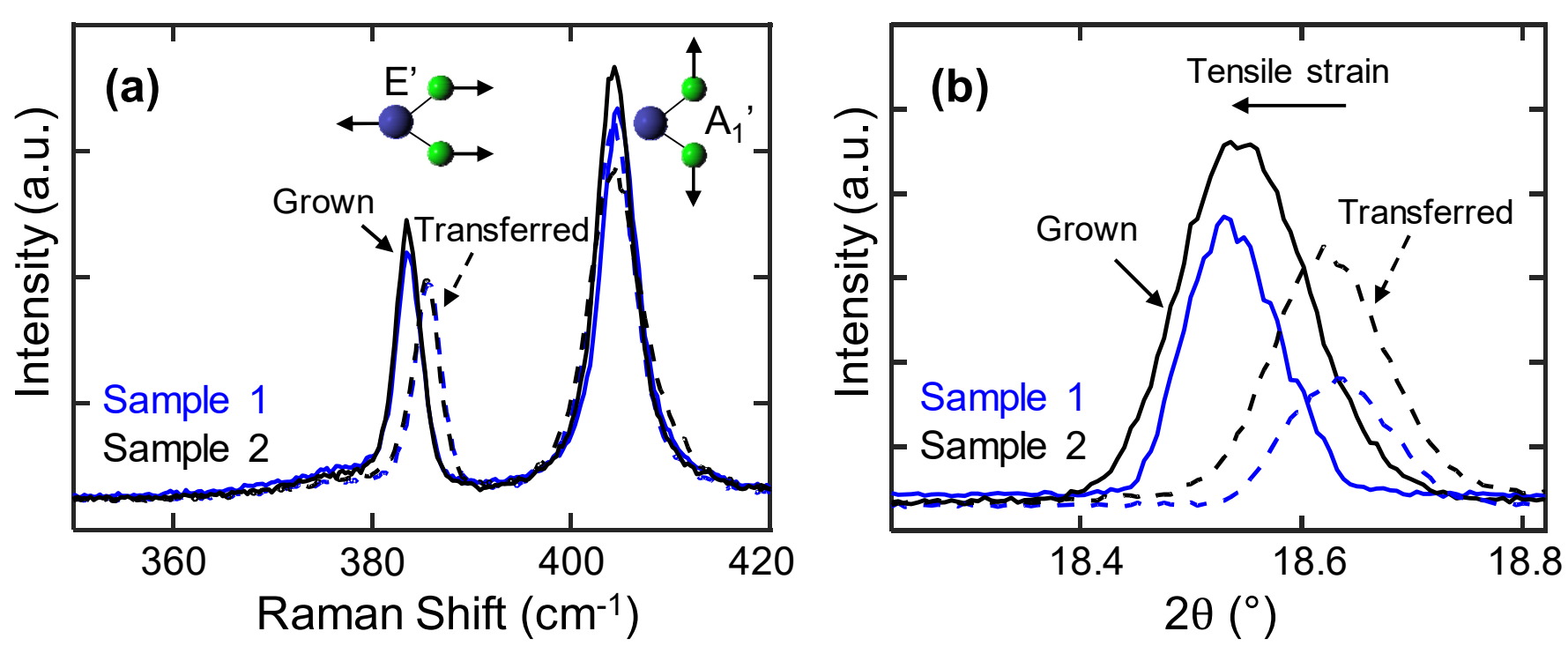

| (c) | ΔPos(E') ($cm^{-1}$) | Strain (Raman) | $d_0$ (Å) | d (Å) | Δd (Å) | Strain (XRD) |
|---|---|---|---|---|---|---|
| **Sample 1** | 2.3 | 0.51% | 2.7327 | 2.7467 | 0.014 | 0.51% |
| **Sample 2** | 1.9 | 0.42% | 2.7431 | 2.7453 | 0.011 | 0.40% |

**Figure S8:** (a) Raman spectra of transferred vs. as-grown $MoS_2$, reveal E' peak red shifts corresponding to ~0.4 to 0.5% tensile strain, depending on the particular growth. (b) Grazing incidence XRD of transferred vs. as-grown monolayer $MoS_2$ in-plane (10) peak. The XRD peak shift also reveals ~0.4 to 0.5% biaxial tensile strain in the as-grown $MoS_2$. Note Raman data are point spectra over < 1 $\mu m^2$ regions (the laser spot size) whereas XRD averages over an effective measurement area of 2.8 mm by the whole length of the sample (7-12 mm). (c) Table shows Raman-based $MoS_2$ strain calculated from change in E' peak position after transfer, and XRD-based $MoS_2$ strain calculated from change in $d$ spacing after transfer ($\Delta d/d_0$). $d_0$ is $d$-spacing after transfer (assumed unstrained), $d$ is before transfer (tensile strained), and $\Delta d$ is the change.

Figure S9 displays Williamson-Hall XRD analysis of bare CVD-grown $MoS_2$, a method where peak broadening due to crystallite size and microstrain is deconvoluted by considering the peak width (FWHM of "as-grown" $MoS_2$ peaks in Figure S8b) as a function of X-ray Bragg angle θ. The Williamson-Hall formula (inset in Figure S9) shows a linear relationship between the (10), (20), and (30) peak FWHM (= $B_{sample}$) multiplied by cos(θ) as a function of sin(θ).[18] The y-intercept of this model gives information on the crystallite size, where K is the Scherrer coefficient, and the slope gives the microstrain within the crystallites, as fit with dashed lines in Figure S9. We find that with K = 1.05,[19] the average crystallite size is 61 ± 4.2 nm, which is smaller than the grain size of the $MoS_2$ film, as the estimated crystallite size is affected by line defects, dislocations, stacking faults, or other disorder. We also estimate the average microstrain in all four samples is 0.084 ± 0.047%, i.e. the distribution of strains across the crystallites.

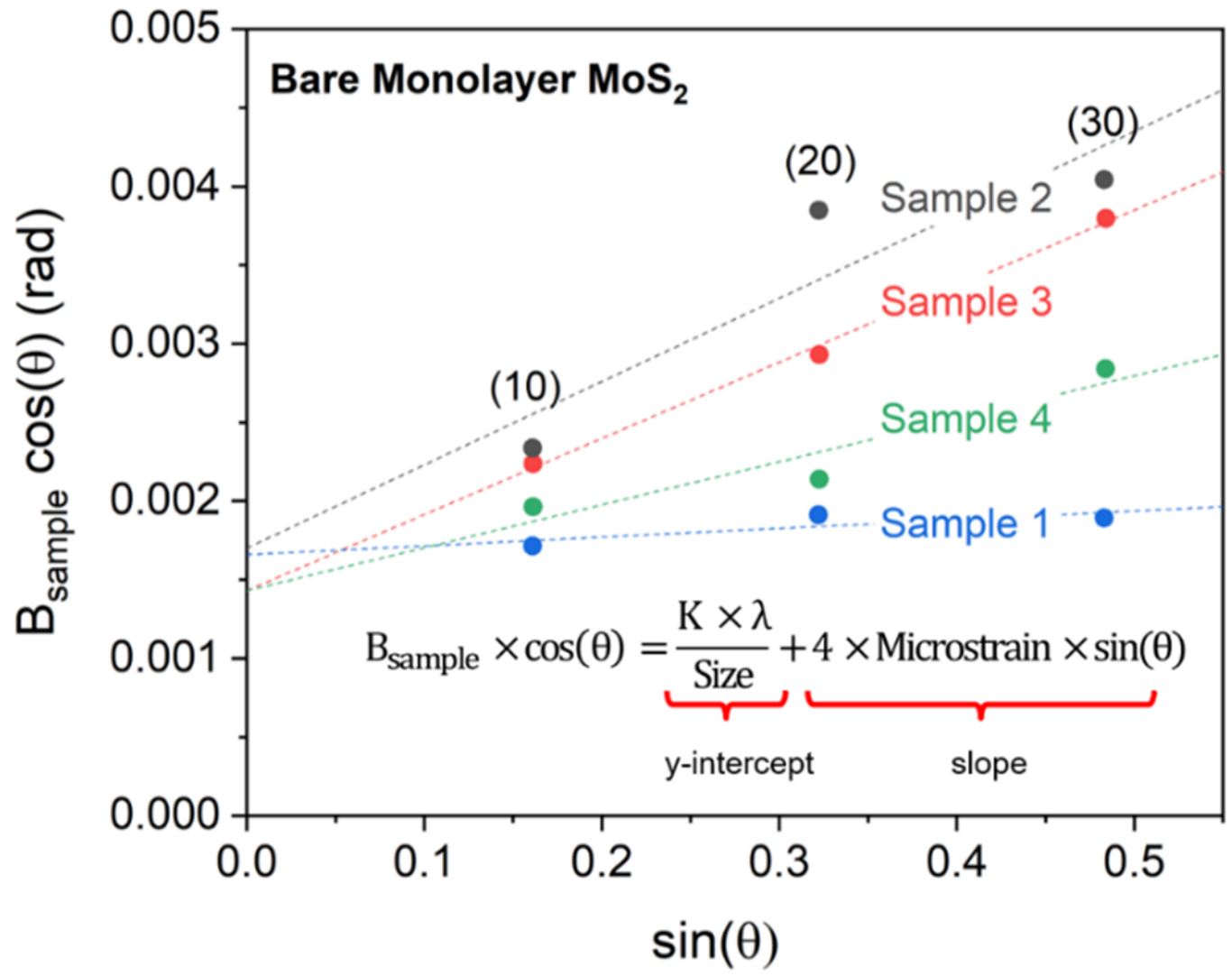

**Figure S9:** Williamson-Hall analysis of four different bare, as-grown monolayer $MoS_2$ growths (on our typical $SiO_2$/Si substrates) using grazing incidence XRD data, where each color represents a different growth. The dashed lines represent linear fits to the scattered data points.

Figure S10 shows the spacing $d$ between (10) rows of $MoS_2$ atoms measured by grazing incidence XRD (calculated from XRD 2θ scans, in main text Figure 5a). Average strain is calculated based on change in lattice spacing relative to the as-grown monolayer $MoS_2$ ($\Delta d/d$), indicating no change in $MoS_2$ strain due to ultrathin metal capping. We note that in the case of contact metals on $MoS_2$, it was not possible to do a Williamson-Hall analysis due to poor signal-to-noise ratio of the (20) peak. As a result, the FWHMs reported in Figure S10 embody the microstrain in addition to other broadening effects such as crystallite size. Overall, the FWHM of the $MoS_2$ (10) XRD peak is similar between bare $MoS_2$ and $MoS_2$ capped with ultrathin metals, indicating no significant changes in $MoS_2$ strain distribution or crystallite size due to metals. However, from transport measurements (main text Figure 4c) we note that some metals, especially Ni, introduce point defects which reduce the electron mobility. (XRD is not sensitive to these point defects.)

| | d spacing (Å) | Strain based on d-spacing | FWHM (°) |
|---|---|---|---|
| Bare $MoS_2$ | 2.7429 | -- | 0.12 |
| Ag on $MoS_2$ | 2.7430 | 0.003% | 0.12 |
| Au on $MoS_2$ | 2.7432 | 0.012% | 0.13 |
| Al on $MoS_2$ | 2.7433 | 0.014% | 0.12 |
| Ni on $MoS_2$ | 2.7439 | 0.037% | 0.11 |

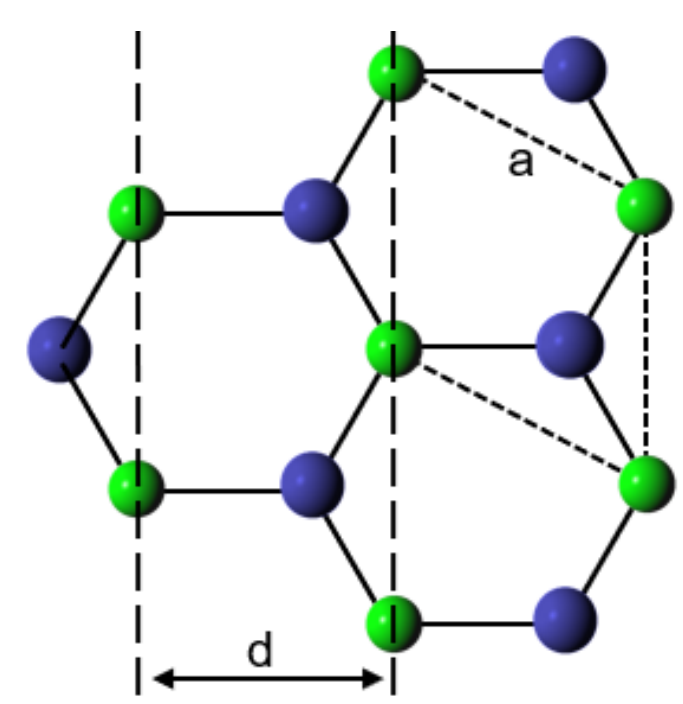

**Figure S10:** Table shows measured $d$-spacing between (10) rows of atoms, calculated strain, and $MoS_2$ (10) XRD peak FWHM values for bare as-grown $MoS_2$ and coated with each non-reacting metal. The strain listed is *relative* to the as-grown bare $MoS_2$ in the top row (itself ~0.4 to 0.5% tensile strained with respect to transferred $MoS_2$, see Figure S8). Figure (right) shows the physical distance $d = 3^{1/2}a/2$ measured by XRD, where $a$ = 3.167 Å (for our tensile strained as-grown $MoS_2$). We note $a_0$ = 3.15 Å is the accepted (unstrained) lattice constant of bulk $MoS_2$.[20, 21]

For XRD measurements, the sample was attached to a six-circle diffractometer and rotated vertically in the χ direction (Figure S11). The sample was covered with a Kapton dome (not shown) and purged with helium gas to improve the signal-to-noise ratio by reducing air scattering and to reduce sample damage. The 14 keV (0.886 Å) incident beam was set to grazing incidence (ω = 0.1°) and the scattered radiation was collimated to 1 mrad by Soller slits and collected by a Vortex point detector. The sample was rocked up and down in the z direction during the measurement to reduce potential beam damage, and each measurement was averaged over 3 exposures to reduce noise.

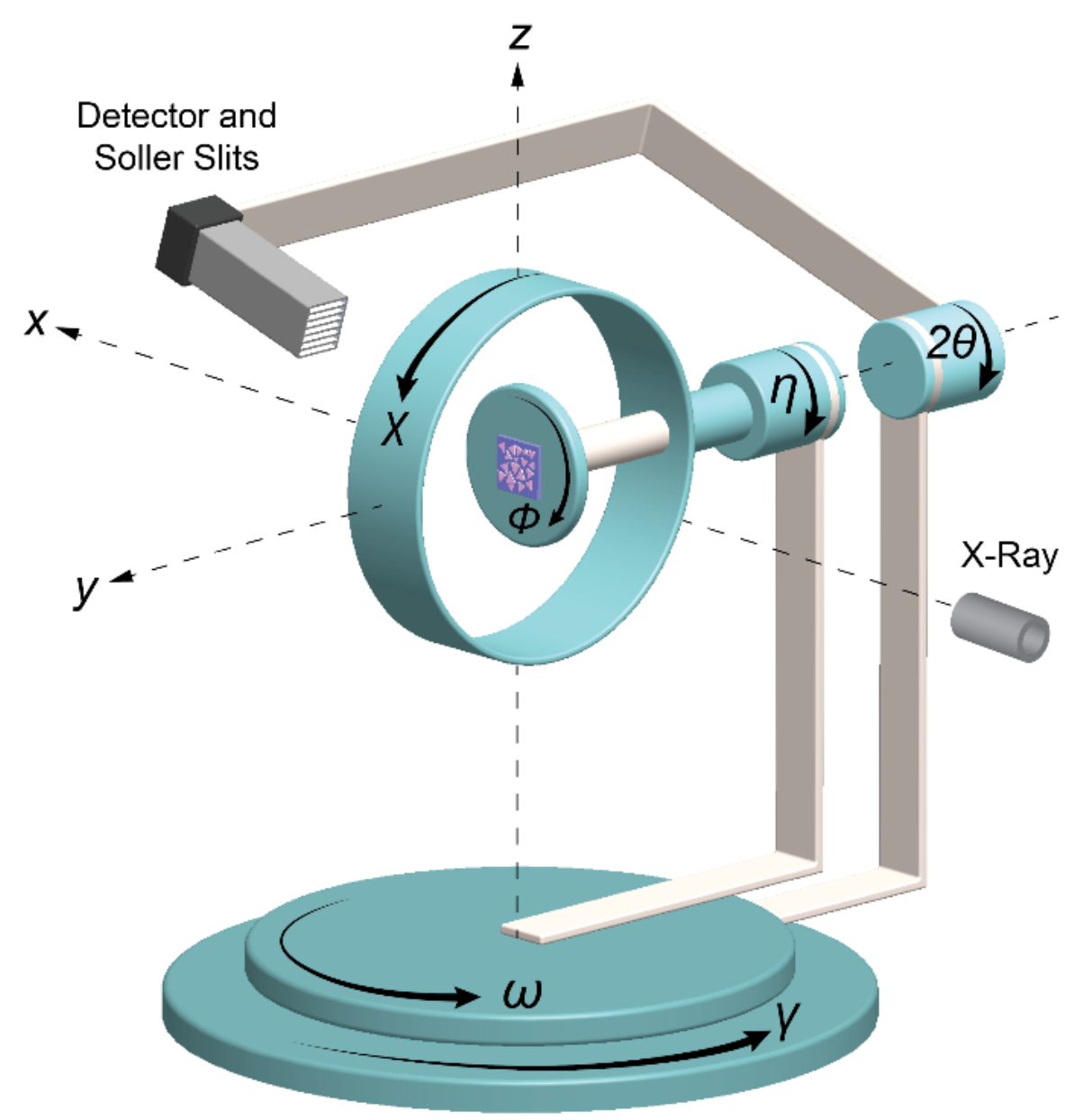

**Figure S11:** X-ray diffraction geometry at the Stanford Synchrotron Radiation Lightsource. The purple square represents the sample, with the small triangles symbolizing polycrystalline $MoS_2$ grains. The labels x, y, and z define the 3D coordinate system, and labels η, ϕ, χ, γ, ω, and 2θ define rotational degrees of freedom. 2θ sweeps the detector arm during measurement and ω sets the incidence angle. Note that η, ϕ, χ, and γ are not used during measurement and, in this configuration, η serves the same purpose as ϕ.

## 7. Supplementary References

1. Nickel, N. H.; Lengsfeld, P.; Sieber, I., Raman spectroscopy of heavily doped polycrystalline silicon thin films. *Phys. Rev. B* **2000,** *61*, 15558-15561.
2. Mignuzzi, S.; Pollard, A. J.; Bonini, N.; Brennan, B.; Gilmore, I. S.; Pimenta, M. A.; Richards, D.; Roy, D., Effect of disorder on Raman scattering of single-layer $MoS_2$. *Phys. Rev. B* **2015,** *91*, 195411.
3. Shi, W.; Zhang, X.; Li, X. L.; Qiao, X. F.; Wu, J. B.; Zhang, J.; Tan, P. H., Phonon Confinement Effect in Two-dimensional Nanocrystallites of Monolayer $MoS_2$ to Probe Phonon Dispersion Trends Away from Brillouin-Zone Center. *Chinese Phys. Lett.* **2016,** *33*, 057801.
4. Crystallography Open Database. http://www.crystallography.net/cod/.
5. Lince, J. R.; Carre, D. J.; Fleischauer, P. D., Schottky-Barrier Formation on a Covalent Semiconductor without Fermi-Level Pinning: The Metal-$MoS_2$(0001) Interface. *Phys. Rev. B* **1987,** *36*, 1647-1656.
6. McGovern, I. T.; Dietz, E.; Rotermund, H. H.; Bradshaw, A. M.; Braun, W.; Radlik, W.; Mcgilp, J. F., Soft-X-Ray Photoemission Spectroscopy of Metal Molybdenum-Disulfide Interfaces. *Surf. Sci.* **1985,** *152*, 1203-1212.
7. Domask, A. C.; Gurunathan, R. L.; Mohney, S. E., Transition Metal-$MoS_2$ Reactions: Review and Thermodynamic Predictions. *J. Electron. Mater.* **2015,** *44*, 4065-4079.

8. Chase, M. W., *NIST-JANAF Thermochemical Tables*; 4th Edition. American Institute of Physics: **1998**.
9. Brewer, L., The Thermodynamic Properties of the Oxides and Their Vaporization Processes. *Chem. Rev.* **1952,** *51*, 1.
10. O'Hare, P. A. G.; Johnson, G. K., Thermochemistry of inorganic sulfur compounds VII. Standard molar enthalpy of formation at 298.15 K, high-temperature enthalpy increments, and other thermodynamic properties to 1100 K of titanium disulfide, $TiS_2$. *J. Chem. Thermodyn.* **1986,** *18*, 189.
11. Thompson, W. T.; Flengas, S. N., Drop Calorimetric Measurements on some Chlorides, Sulfides, and Binary Melts. *Can. J. Chem.* **1971,** *49*, 1550.
12. Nakahara, J. F. The electronic structure and thermodynamics of scandium monosulfide. Iowa State University, Retrospective Theses and Dissertations, 1984.
13. Gschneider, K. A. J.; Kippenhan, N. *Thermochemistry of the rare earth carbides, nitrides, and sulfides for steelmaking*; U.S. Department of Energy Office of Scientific and Technical Information, 1971.
14. Smithe, K. K. H.; Suryavanshi, S. V.; Muñoz Rojo, M.; Tedjarati, A. D.; Pop, E., Low variability in synthetic monolayer $MoS_2$ devices. *ACS Nano* **2017,** *11*, 8456-8463.
15. Dismukes, J. P.; White, J. G., The Preparation, Properties, and Crystal Structures of Some Scandium Sulfides in the Range $Sc_2S_3$-ScS. *Inorganic Chemistry* **1964,** *3*, 1220-1228.
16. Vaziri, S.; Yalon, E.; Munoz Rojo, M.; Suryavanshi, S. V.; Zhang, H.; McClellan, C. J.; Bailey, C. S.; Smithe, K. K. H.; Gabourie, A. J.; Chen, V.; Deshmukh, S.; Bendersky, L.; Davydov, A. V.; Pop, E., Ultrahigh thermal isolation across heterogeneously layered two-dimensional materials. *Science Advances* **2019,** *5*, eaax1325.
17. Li, H.; Contryman, A. W.; Qian, X. F.; Ardakani, S. M.; Gong, Y. J.; Wang, X. L.; Weisse, J. M.; Lee, C. H.; Zhao, J. H.; Ajayan, P. M.; Li, J.; Manoharan, H. C.; Zheng, X. L., Optoelectronic crystal of artificial atoms in strain-textured molybdenum disulphide. *Nat. Commun.* **2015,** *6*, 7381.
18. Warren, B. E., *X-Ray Diffraction*; Dover Publications: New York, **1990**.
19. Lele, S.; Anantharaman, T. R., Influence of Crystallite Shape on Particle Size Broadening of Debye-Scherrer Reflections. *Proc. Indian Acad. Sci.* **1966,** *64*, 261-274.
20. Dickinson, R. G.; Pauling, L., The Crystal Structure of Molybdenite. *J. Am. Chem. Soc.* **1923,** *45*, 1466-1471.
21. Wakabayashi, N.; Smith, H. G.; Nicklow, R. M., Lattice dynamics of hexagonal $MoS_2$ studied by neutron scattering. *Phys. Rev. B* **1975,** *12*, 659-663.